**Root Cause Analysis of Radiation Oncology Incidents Using Large Language Models**

Yuntao Wang, MS[1], Mariluz De Ornelas, PhD[1], Matthew T. Studenski, PhD[1], Elizabeth Bossart, PhD[1], Siamak P. Najad-Davarani, PhD[1], Yunze Yang, PhD[1*]

[1]Department of Radiation Oncology, University of Miami, FL 33136, USA

*Corresponding Author

Yunze Yang, PhD, Assistant Professor of Radiation Oncology

Department of Radiation Oncology, Sylvester Comprehensive Cancer Center, University of Miami, 1475 NW 12th Ave, Miami, FL 33136

E-mail: yxy971@med.miami.edu

Author Responsible for Statistical Analysis Name & Email Address

Yunze Yang, PhD, yxy971@med.miami.edu

**Abstract**

Purpose

To evaluate the reasoning capabilities of large language models (LLMs) in performing root cause analysis (RCA) of radiation oncology incidents using narrative reports from the Radiation Oncology Incident Learning System (RO-ILS), and to assess their potential utility in supporting patient safety efforts.

Methods and Materials

We prompted four state-of-the-art LLMs, Gemini 2.5 Pro, GPT-4o, o3, and Grok 3, with the "Background and Incident Overview" sections from 19 publicly available RO-ILS cases. Each model was instructed to perform RCA and generate root causes, lessons learned, and suggested actions using a standardized prompt based on AAPM RCA guidelines. Model outputs were evaluated using a combination of objective semantic similarity metrics (cosine similarity via Sentence Transformer), semi-subjective assessments (precision, recall, F1-score, accuracy, hallucination rate and performance criteria including relevance, comprehensiveness, quality of justification and quality of solution), and subjective ratings (reasoning quality and overall performance) by five board-certified medical physicists.

Results

LLMs demonstrated satisfactory performance across evaluation metrics. GPT-4o achieved the highest cosine similarity (0.831), and Gemini 2.5 Pro had the highest recall (0.799) and accuracy (0.918). All models exhibited some degree of hallucination, ranging from 11% to 61%. Gemini 2.5 Pro, which outperformed all other models across performance evaluation criteria, received an overall performance rating of 4.8 out of 5 from expert reviewers. Statistically significant differences were observed among models in accuracy, hallucination rate, and subjective ratings ($p < 0.05$).

Conclusion

LLMs delivered promising results as assistive tools for RCA in radiation oncology, with the ability to generate relevant and accurate analyses aligned with expert expectations. LLMs may support incident analysis and contribute to quality improvement efforts to advance patient safety in clinical radiation oncology practice.

## Introduction

The integration of generative artificial intelligence (GAI), particularly Large Language Models (LLMs), has rapidly expanded across the medical landscape, offering transformative solutions for tasks ranging from administrative automation to complex clinical decision support[1-3]. The transition of LLMs from advanced pattern recognition to architectures capable of emergent reasoning represents a fundamental breakthrough in GAI, driving a transformative leap in clinical medicine[4-6]. This progress stems from immense model scale and architectural innovation[7]. Models with trillions of parameters capture complex relationships from vast datasets[8], while architectures like Mixture-of-Experts (MoE) improve efficiency by selectively activating specialized sub-networks for specific tasks[9]. This is also cultivated through multiple approaches in training and prompting[10]. Training strategies, such as Supervised Fine-Tuning (SFT)[11] and Reinforcement Learning from Human Feedback (RLHF)[12, 13], fundamentally align the model's logic with human intent and coherence. In addition, dynamic prompting strategies like Chain-of-Thought (CoT)[14] and Tree-of-Thought (ToT)[15] are employed to structure the model's deliberation, guiding it to break down complex problems and explore solutions step-by-step[16].

Such progression of reasoning capabilities is evidenced by their improving performance on a variety of complex tasks[17], including natural language understanding, code generation[18], and problem-solving[19, 20]. The development of diverse benchmarks such as MMLU (Massive Multitask Language Understanding)[19], GPQA (A Challenging Benchmark for Advanced Reasoning)[21], and various coding and commonsense reasoning tests (e.g., HumanEval, HellaSwag) have been crucial in quantifying these advancements[19]. Models' mathematical reasoning is evaluated on benchmarks ranging from grade-school problems (GSM8K)[20] to advanced competition math (MATH)[19]. Pushing this frontier, a recent Gemini model demonstrated capabilities competitive with the brightest human minds, achieving a gold-medal standard in the International Mathematical Olympiad (IMO) competition[22]. However, evaluating the reasoning capabilities of LLMs remains an ongoing challenge. Issues such as inherent biases learned from vast training datasets, the potential for "hallucination" or the generation of incorrect information, and the reproducibility and robustness of their reasoning processes are active areas of research[23, 24]. The sensitivity of these models to different prompting strategies also complicates efforts to ensure their reliability and trustworthiness, a critical concern in safety-critical applications.

Radiation oncology is a highly complex medical discipline that integrates advanced technology, intricate clinical workflows, and multidisciplinary collaboration. The treatment process spans multiple stages, from imaging and contouring to treatment planning, quality assurance, and daily delivery, creating numerous interfaces where systemic failures or human errors can occur. Given the high doses of radiation involved, ensuring patient safety and treatment quality is paramount[25, 26]. Radiation Oncology Incident Learning Systems (RO-ILS), managed by American Society for Radiation Oncology (ASTRO), enable confidential reporting, collection, and analysis of errors, near misses, and unsafe conditions with narrative texts[27]. The primary goal of RO-ILS is to foster a culture of safety through shared learning from these incidents to prevent future occurrences[28, 29]. A critical component of analyzing reported incidents is Root Cause Analysis (RCA)[30, 31]. Current RCA methodologies in radiation oncology, often guided by American Association of Physicists in Medicine (AAPM) recommendations and adapted from broader healthcare and industrial safety practices, typically involve a systematic process. This process includes forming a multidisciplinary team, gathering data surrounding the event, creating a sequence or timeline of events, repeatedly asking "whys" to drill down to fundamental causes, identifying contributing factors, and formulating corrective and preventive actions. This structured, human-driven analytical process is essential for understanding the complex interplay of factors that can lead to incidents in the radiation therapy workflow.

The intersection of rapidly advancing reasoning capabilities of LLMs and the critical need for robust safety analyses in specialized domains like radiation oncology presents a compelling area of investigation. This paper seeks to perform RCA of radiation oncology incidents using LLMs and to evaluate their reasoning capabilities within this highly specific and safety-critical domain. This task is uniquely challenging as it demands not only a nuanced understanding of domain-specific knowledge, encompassing clinical procedures, medical physics, radiation therapy equipment, and quality assurance protocols, but also sophisticated reasoning abilities. These abilities include identifying causal relationships, understanding complex procedural workflows, and synthesizing information from incident reports to pinpoint underlying systemic failures rather than just superficial errors. By evaluating LLMs' ability to perform RCA in radiation oncology, this study aims to illuminate their potential utility in clinical workflow optimization and quality improvement, and investigate the current limitations in supporting patient safety efforts.

## Methods and Materials

### RO-ILS Study Cases

Figure 1 outlines the schematic diagram of the study design. We utilized a dataset of nineteen (n = 19) publicly available study cases from the RO-ILS prior to May 2025[32]. A representative case is provided in Supplementary Material 1. These cases were selected to represent a variety of incident types and complexities encountered in clinical practice. Each published RO-ILS case typically includes comprehensive documentation comprising:

- Background and Incident Overview: A detailed narrative of the context, sequence of events leading to the incident, and the nature of the incident itself.
- Root Cause(s) and/or Contributing Factors: Expert-identified primary and secondary factors that led to the event.
- Lessons Learned: Key takeaways and insights derived from the analysis of the incident.
- Suggestions and Actions: Recommended corrective and preventative measures to mitigate future occurrences.

The "Background and Incident Overview" section served as the input for each LLM, while the remaining sections of the reports were used as the ground truth reference for evaluation.

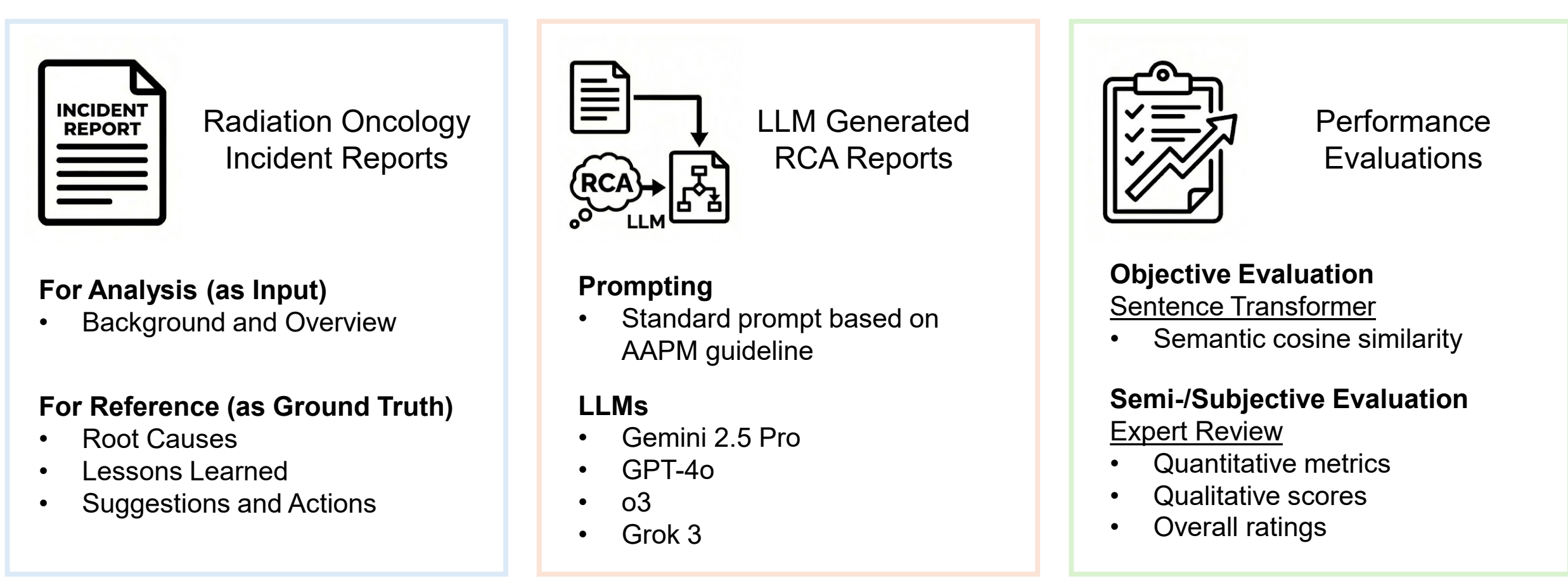

Figure 1 Schematic overview of the study design. (Left) Radiation oncology incident reports from 19 cases were utilized for this study. The "Background and Overview" sections served as inputs for the LLM, while the original sections containing root causes, lessons learned, suggestions, and actions served as the ground truth. (Center) LLMs were utilized to process the event data and generate structured RCA reports, using prompts derived from AAPM guidelines. (Right) Performance evaluation assessed the model outputs using objective, semi-subjective, and subjective metrics for a comprehensive quantification.

**LLM Reasoning Testing**

We tested four state-of-the-art LLMs: Gemini 2.5 Pro (Google)[33], GPT-4o (OpenAI)[34], o3 (OpenAI), and Grok 3 (Grok). For each of the 19 ROI-LS cases, the "Background and Incident Overview" section was provided as input to each LLM. The models were then prompted to perform an RCA and generate associated recommendations using the following standardized prompt in Table 1:

Table 1 Standardized prompt used for RCA of RO-ILS cases

| **Standardized prompt sentences:** |
| --- |
| *"Please make a standardized root cause analysis (RCA) for this radiation oncology medical event/incident according to the AAPM standard:*<br>*Simple Framework for RCA includes at least:*<br>*• Chronological sequence– Diagram the flow of events leading up to the incident (including the three "whys")*<br>*• Cause and Effect Diagramming– Identify the conditions that resulted in the adverse event or close call*<br>*• Causal Statements– Develop root cause and contributing factor statements, actions, and outcome*<br>*After that, please summarize in short bullet points (as many points as needed) for a section of Lessons Learned from this incident, and a section of Suggestions and Actions."* |

The LLM-generated outputs for "Root Cause(s)/Contributing Factors," "Lessons Learned," and "Suggestions and Actions" were collected for subsequent evaluation.

**Model Evaluation**

The performance of each LLM was assessed through a combination of objective, semi-subjective and subjective measures.

*Objective Evaluation:*

We employed text semantic similarity metrics to quantify the overlap between LLM-generated content and the reference text from the corresponding RO-ILS case study. Specifically, cosine similarity was calculated using Sentence Transformers ("all-mpnet-base-v2" model)[35] for the "Root Cause(s)/Contributing Factors," "Lessons Learned/Suggestions and Actions" sections and the combination of both sections (referred as the overall texts).

*Semi-Subjective Evaluation:*

A panel of five board-certified medical physicists, each with extensive experience in clinical radiation oncology and incident analysis, participated in this phase. Each of the 19 cases, along with the corresponding LLM outputs, was independently evaluated by two physicists to ensure inter-rater reliability. The physicists were blinded to the specific LLM generating each output during their primary review. Following the initial review, scoring discrepancies were discussed and re-evaluated if the difference in objective metrics (precision, recall, F1-score, accuracy) exceeded 0.5, or if the Likert scale scores differed by more than 2. The following aspects were evaluated:

- Root Cause/Contributing Factors Identification: Precision, recall, and F1-score were calculated based on the physicists' judgment of whether the LLM correctly identified the key root causes and contributing factors present in the ground truth RO-ILS report (from 0 to 1, on numerical values). In addition, accuracy was calculated based on the physicist's assessment of the correctness of the identified root causes and contributing factors in the LLM responses (from 0 to 1, on numerical values). Precision, recall, F1-score and accuracy are defined in the following equations:

$$\text{Precision} = \frac{\text{\# of LLM identified items in Case Study}}{\text{\# of LLM identified items}}$$

$$\text{Recall} = \frac{\text{\# of LLM identified items in Case Study}}{\text{\# of total items identified in Case Study}}$$

$$\text{F1} - \text{score} = 2 \times \frac{\text{Precision} \times \text{Recall}}{\text{Precision} + \text{Recall}}$$

$$\text{Accuracy} = \frac{\text{\# of LLM identified items are correct}}{\text{\# of LLM identified items}}$$

- Hallucination: Evaluates if the LLM's response is factually correct and identifies if the LLM's response includes fabricated, irrelevant, or inaccurate information not supported by the incident description (on binary yes/no: 1=fabricated information identified, 0=no fabricated information identified).
- Relevance: Evaluates whether the model's entire response is pertinent to the specific case study (on a 5-point Likert scale).
- Comprehensiveness: Assesses the comprehensiveness of the identified root causes and actions provided by the LLM (on a 5-point Likert scale).
- Quality of Justification: Assesses the model's ability to provide clear, logical, and coherent explanations for the identified root causes (on a 5-point Likert scale).
- Quality of Solutions: Assesses the model's ability to provide reasonable, actionable, and relevant suggestions and actions (on a 5-point Likert scale).

All assessments were evaluated based on the experience and domain knowledge of medical physicists. We referred to these evaluations as semi-subjective.

*Subjective Evaluation:*

Following the detailed semi-subjective assessment, the same physicists were asked to provide an overall subjective score for each LLM's output per case. This included:

- Reasoning Capability: A rating on a 5-point Likert scale (1=Very Poor, 5=Excellent) reflecting the perceived quality of the LLM's analytical and inferential reasoning in performing the RCA.
- Overall Performance: A rating on a 5-point Likert scale (1=Very Poor, 5=Excellent) summarizing the overall utility and quality of the LLM's complete response for the given case.

The table and detailed rubric used by medical physicists for their evaluations is provided in Supplementary Material 2.

**Statistical Analysis**

All quantitative data, including objective metric scores, and subjective Likert scale ratings, were quantified using descriptive statistics across cases for each LLM. Friedman's test was used to determine if there are statistically significant differences in performance between the LLMs. A *p*-value of $< 0.05$ was considered statistically significant. All statistical analyses were performed using Python 3.13 and SciPy v1.16.1 package.

**Results**

RCA reports from the four LLM models were generated following the prompting process. Four representative reports, generated using distinct LLM models, are presented in Supplementary Material 3. Figure 2 presents box plots of the semantic cosine similarity scores for the four models. For the overall texts, the cosine similarity scores for Gemini 2.5 Pro, o3, GPT-4o and Grok 3 were 0.804 ± 0.079 [mean ± std], 0.793 ± 0.067, 0.831 ± 0.051, and 0.804 ± 0.074, respectively. A similar pattern of performance was observed for the "Contributing Factors / Root Cause" section, and " Lessons Learned / Suggestions and Actions" section. Generally, model GPT-4o demonstrated highest cosine similarity score with the overall reference texts and section "Lessons Learned / Suggestions and Actions", while Gemini 2.5 Pro achieved highest similarity in the "Root Causes / Contributing Factors" section, as detailed in Table 2. Friedman's test showed statistically significant differences in model performance for "Lessons Learned / Suggestions and Actions" ($p = 0.001$), and "Root Causes / Contributing Factors" sections ($p = 0.010$), but not in their combined texts ($p = 0.126$) Detailed similarity scores for each case studies were presented in Supplementary Figure 1.

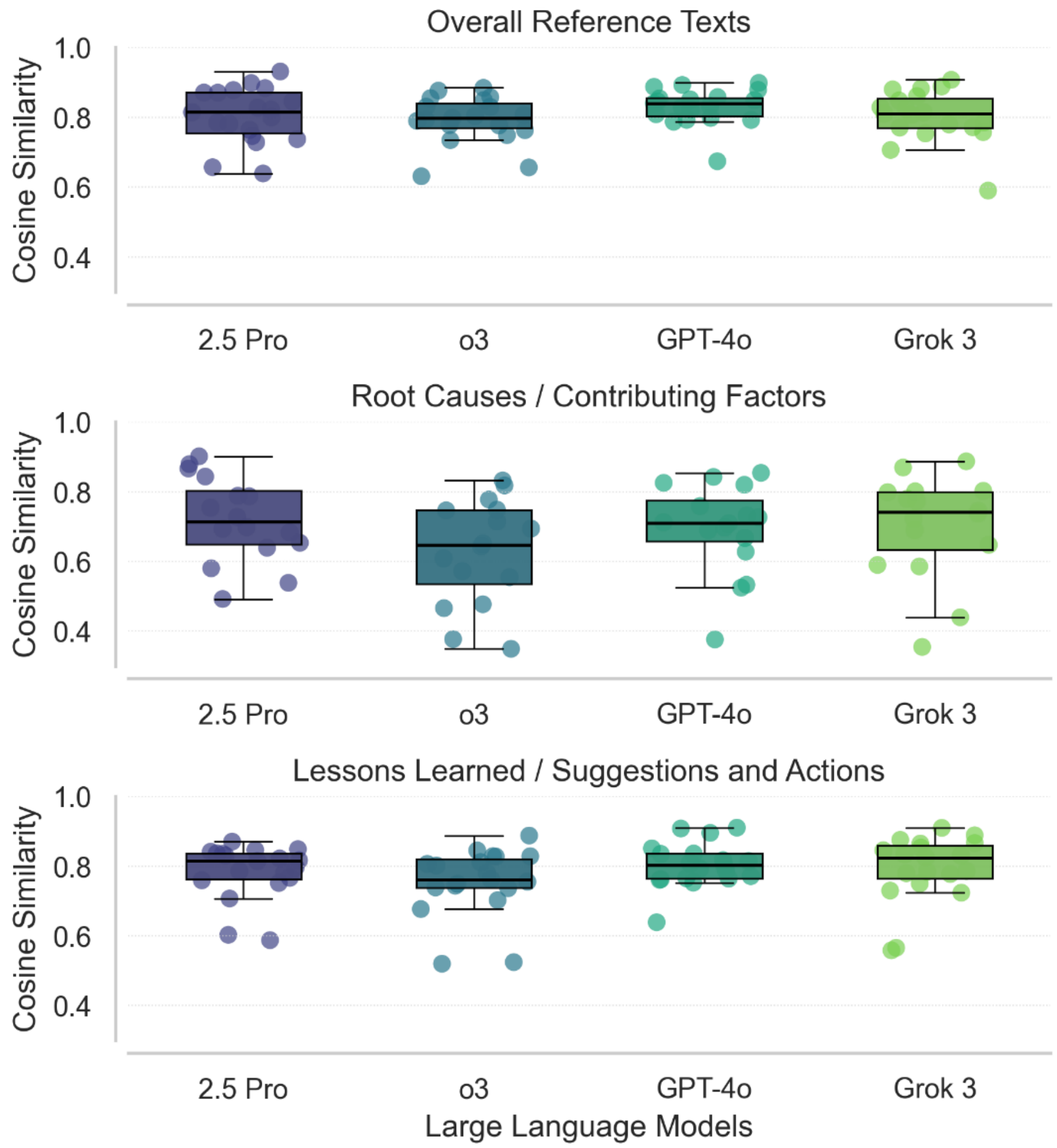

Figure 2 Box plots of cosine similarity scores for LLM outputs compared against three sections of the reference texts: the combined sections (top), "Root Causes / Contributing Factors" (middle), and "Lessons Learned / Suggestions and Actions" (bottom).

Table 2 Cosine similarity scores of four LLMs for different sections

| Sections | LLM Models | | | | *P*-value |
|---|---|---|---|---|---|
| | 2.5 Pro | o3 | GPT-4o | Grok 3 | |
| Overall Reference Texts | 0.804 ± 0.079 | 0.793 ± 0.067 | 0.831 ± 0.051 | 0.804 ± 0.074 | 0.126 |
| "Root Cause(s) / Contributing Factors" | 0.711 ± 0.12 | 0.614 ± 0.15 | 0.683 ± 0.126 | 0.690 ± 0.143 | 0.010 |
| "Lessons Learned / Suggestions and Actions" | 0.782 ± 0.079 | 0.751 ± 0.093 | 0.803 ± 0.062 | 0.793 ± 0.099 | 0.001 |

Table 3 summarizes the performance metrics results from physicists' evaluation. In the task of identifying root causes and contributing factors, Gemini 2.5 Pro achieved a mean precision of 0.705 ± 0.297, recall of 0.799 ± 0.256, and an F1-score of 0.727 ± 0.268. While aforementioned metrics were based on the root causes documented in the case study report, the accuracy of root causes identified by Gemini 2.5 Pro was 0.918 ± 0.19 when judged by the expert physicist panel. LLMs of o3, GPT-4o and Grok 3 displayed a similar trend (Figure 3a, Figure 3c and Table 3). Among all four models, Gemini 2.5 Pro led in all four performance metrics. Statistical analysis revealed a significant difference in accuracy across the evaluated models ($p = 0.002$). No significant differences were found for the other performance metrics (Table 3).

Instances of hallucination, defined as fabricated or inaccurate information, were observed in 11% of cases for Gemini 2.5 Pro, 61% for o3, and 32% for GPT-4o, and 29% for Grok 3 respectively on average (Figure 3b, Table 3). The hallucination score differed statistically significantly among the models ($p < 0.001$).

Table 3 Performance metrics and hallucination evaluation for the four LLMs.

| Performance Metrics | Model | | | | *P*-value |
|---|---|---|---|---|---|
| | 2.5 Pro | o3 | GPT-4o | Grok 3 | |
| Precision | 0.705 ± 0.297 | 0.658 ± 0.273 | 0.698 ± 0.278 | 0.614 ± 0.309 | 0.091 |
| Accuracy | 0.918 ± 0.190 | 0.768 ± 0.285 | 0.880 ± 0.238 | 0.816 ± 0.301 | 0.002 |
| Recall | 0.799 ± 0.256 | 0.689 ± 0.275 | 0.752 ± 0.263 | 0.717 ± 0.298 | 0.113 |
| F1 Score | 0.727 ± 0.268 | 0.654 ± 0.252 | 0.694 ± 0.231 | 0.644 ± 0.286 | 0.225 |
| Hallucination | 0.105 ± 0.311 | 0.605 ± 0.495 | 0.316 ± 0.471 | 0.289 ± 0.460 | < 0.001 |

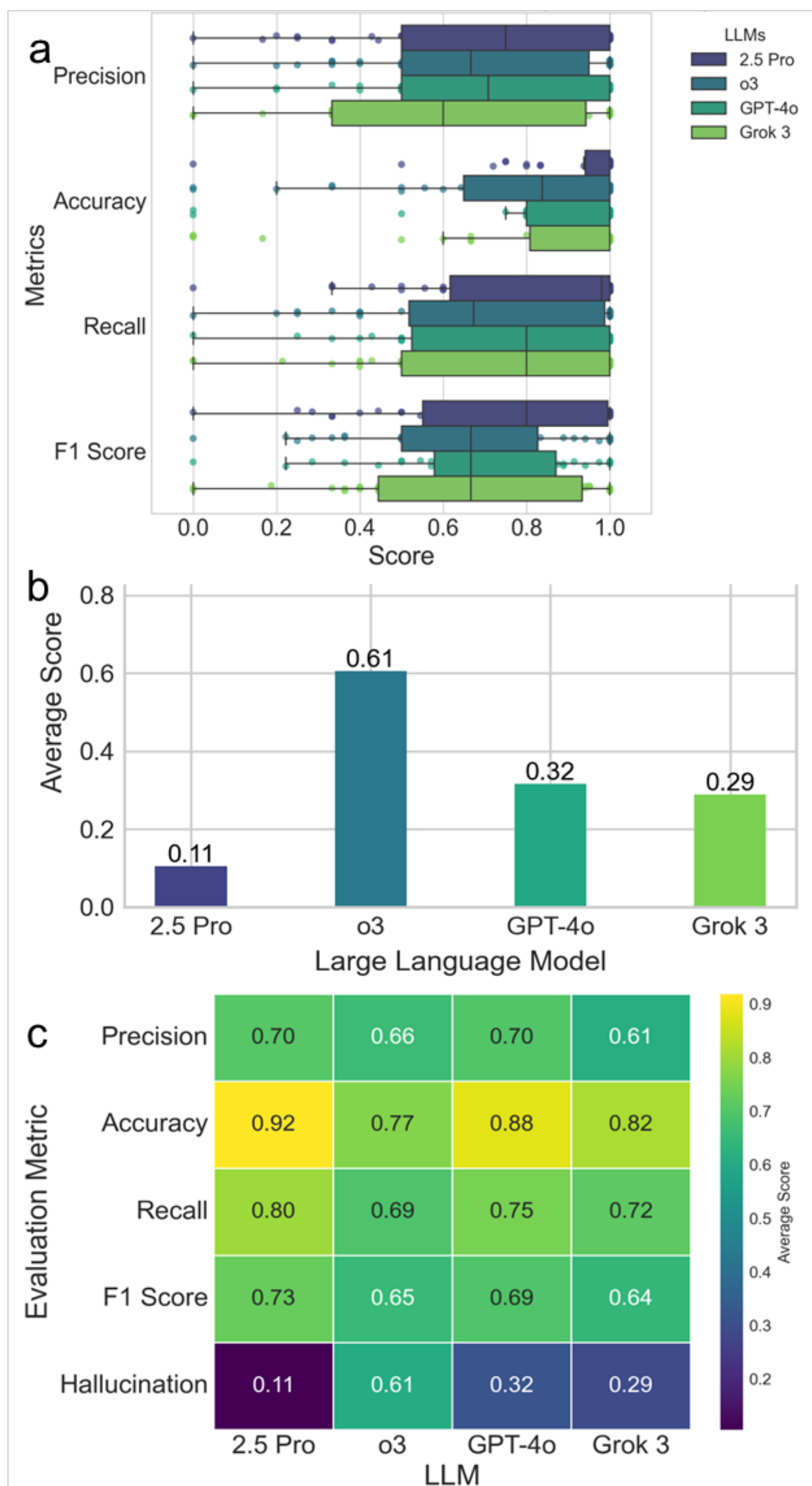

Figure 3 Performance and hallucination analysis for the four evaluated LLMs. (a) Box plot of the distribution of performance metrics. (b) Bar plot of the mean hallucination scores. (c) Heatmap of the relationship between mean performance metrics and hallucination scores.

All models demonstrated satisfactory performance across the evaluated criteria, including Relevance, Comprehensiveness, Quality of Justification, Quality of Solutions, as well as in the subjective assessments of Logical Reasoning and Overall Performance (Figure 4 and Table 4). Among the four models, Gemini 2.5 Pro consistently outperformed the others, with scores of $4.789 \pm 0.413$ for Relevance, $4.921 \pm 0.273$ for Comprehensiveness, $4.842 \pm 0.37$ for Quality of Justification, and $4.789 \pm 0.413$ for Quality of Solutions. It also achieved an average score of 4.8 out of 5 in Logical Reasoning and Overall Performance as rated by expert evaluators. All performance criteria and overall ratings were found to be statistically significantly different among models ($p < 0.001$).

Table 4 Model performances in Relevance, Comprehensiveness, Quality of Justification, Quality of Solutions and subjective ratings of Logical Reasoning Score and Overall Performance Score of four LLMs.

| Performance Metrics | Model | | | | *P*-value |
|---|---|---|---|---|---|
| | 2.5 Pro | o3 | GPT-4o | Grok 3 | |
| Relevance | 4.789 ± 0.413 | 4.211 ± 1.018 | 4.447 ± 0.921 | 4.500 ± 0.952 | < 0.001 |
| Comprehensiveness | 4.921 ± 0.273 | 4.105 ± 0.924 | 4.421 ± 0.976 | 4.395 ± 1.079 | < 0.001 |
| Quality of Justification | 4.842 ± 0.370 | 4.105 ± 1.060 | 4.316 ± 1.016 | 4.342 ± 1.097 | < 0.001 |
| Quality of Solutions | 4.789 ± 0.413 | 4.079 ± 0.997 | 4.421 ± 0.948 | 4.500 ± 0.923 | < 0.001 |
| Logical Reasoning Score | 4.789 ± 0.413 | 4.079 ± 0.997 | 4.368 ± 0.913 | 4.421 ± 0.976 | < 0.001 |
| Overall Performance Score | 4.816 ± 0.393 | 3.763 ± 1.173 | 4.263 ± 0.978 | 4.263 ± 1.131 | < 0.001 |

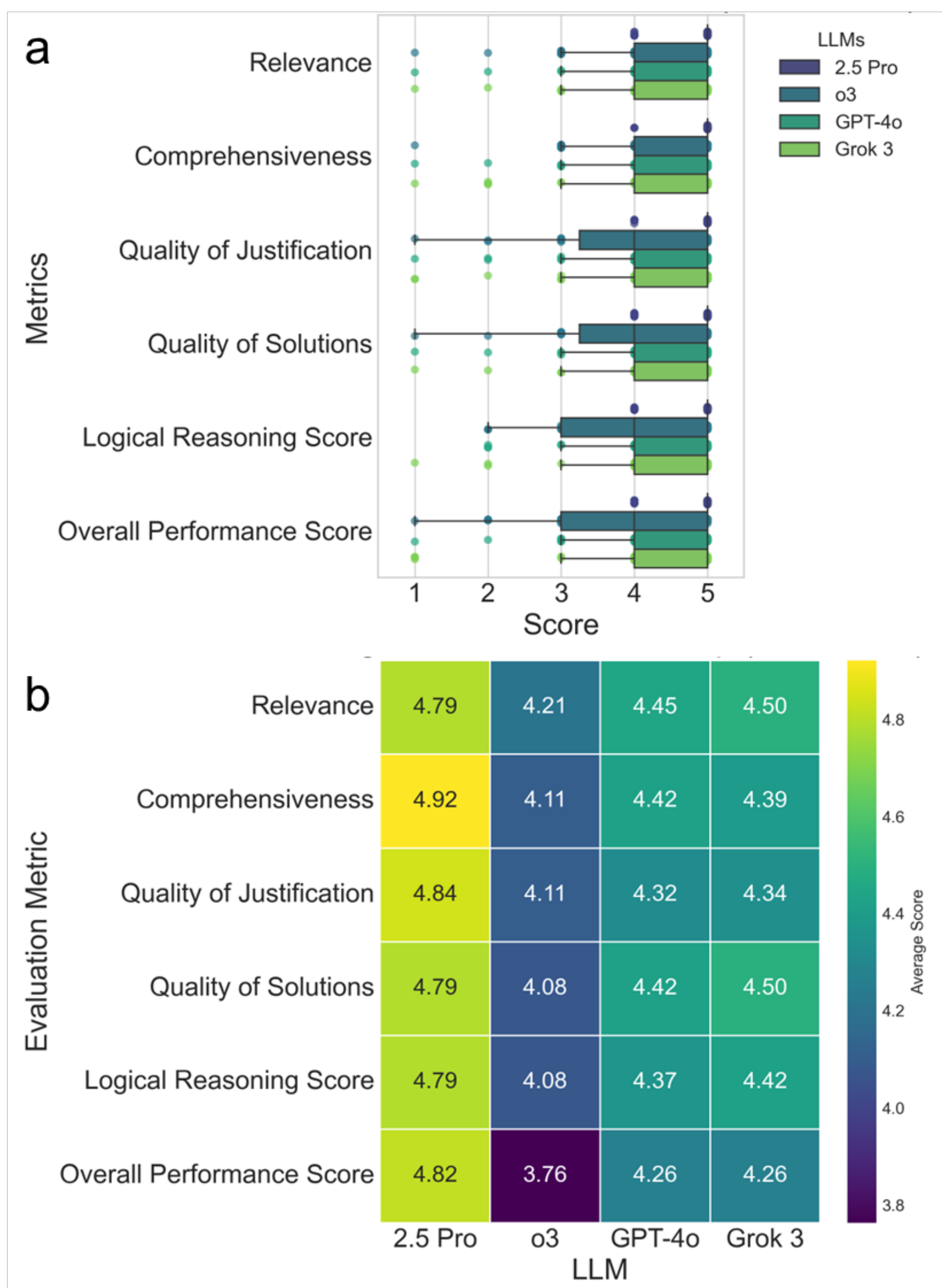

Figure 4 Evaluation of performance for four LLMs. (a) Box plot of the score distribution for six key metrics: Relevance, Comprehensiveness, Quality of Justification, Quality of Solutions, Logical Reasoning, and Overall Performance. (b) Heatmap of the mean score of the six evaluation metrics.

## Discussion

Our study demonstrates the promising, albeit still developing, reasoning capabilities of leading LLMs when applied to the complex task of RCA for radiation oncology incidents. The findings indicate that LLM models can generate RCAs that exhibit considerable alignment with expert-derived analyses. Specifically, the LLMs achieved an average semantic similarity score as high as 0.831 when comparing their generated root causes, lessons learned, and suggested actions against the gold-standard reference texts. Additionally, the best-performing LLM in our evaluation achieved an accuracy of 0.918 in identifying root causes with 11% identified hallucination events and received an overall performance rating of 4.8 out of 5, as assessed by experienced medical physicists. This suggests an emerging potential for LLMs to assist in the initial stages of incident analysis, potentially streamlining the process for human experts.

Despite their advanced capabilities, we observed persistent presence of hallucinations across all evaluated LLMs in this task. The 11% hallucination rate observed in the top-performing model, while not negligible, may still be acceptable for non-patient-facing clinical applications such as RCA of safety incidents, as explored in this study. In this context, LLMs can assist by generating initial drafts or highlighting potential contributing factors that might be overlooked by human reviewers. Expert physicists or oncologists would then review, verify, and refine the output before conclusions are drawn. Thus, the LLM serves as an assistive tool to augment, not replace, expert judgment, mitigating hallucination risks while leveraging its analytical strengths.

Interestingly, we observed a case in which one of the four LLMs refused to generate a root cause analysis, despite being strongly prompted to do so (Supplementary Material 4). Upon expert review, it was confirmed that the incident did not involve any patient harm or workflow failure and merely represented a routine alert during daily image-guided treatment. Although the model's response failed all predefined evaluation metrics, it ultimately offered the true interpretation of the scenario. All the other three tried to fit the data to the RCA structure by making up a story to fill in the gaps. This observation suggests that the model may possess the capacity to recognize the absence of genuine causal relationships, demonstrating contextual understanding that is resilient to conversational prompting bias.

LLMs have shown expert level in tasks related to medical question answering[36, 37] and reasoning[38]. They have also been increasingly integrated into radiation oncology[39] for tasks such as language-assisted contouring[40, 41] and treatment planning[42, 43], clinical information summarization and retrieval[44], clinical question and inquiry response[45, 46], and education[47]. It has also been applied to identify the topics of radiation oncology incidents[48]. These valuable applications primarily capitalize on the text-processing and understanding strengths inherent to LLMs. In contrast, our research focuses on leveraging the rapidly evolving reasoning capabilities of these models, aiming to move beyond text-based automation to address more complex clinical applications. The task of performing RCA in radiation oncology incidents also presents distinct challenges compared to general reasoning tasks often evaluated by standard LLM benchmarks. While general benchmarks test logical deduction, common sense, or factual recall, RCA in this domain requires a deep understanding of highly specific medical and technical knowledge, intricate clinical workflows, and human factors unique to radiation therapy. The incidents are often complex, multi-causal, and require not just logical deduction but also abductive reasoning to infer the most plausible underlying causes from incomplete narratives. Furthermore, unlike many well-defined reasoning tasks, RCA in this context is an inherently open-ended problem. While structured by frameworks like the AAPM recommendations, the identification of specific causes and the formulation of effective "suggestions and actions" demand a degree of contextual understanding and real-world problem-solving that goes beyond simple pattern matching. The ability of the tested LLMs to produce relevant outputs in this specialized, logic-driven yet open-ended task suggests an encouraging capacity for detailed, domain-aware reasoning of LLMs. Evaluating such sophisticated reasoning capabilities accurately is an ongoing challenge. In this study, we employed a dual approach, combining objective text similarity metrics (Sentence Transformer) with comprehensive semi-subjective and subjective evaluations by experienced medical physicists. While objective metrics offer quantifiable measures of textual overlap with ground truth reports, they may not fully capture the semantic correctness, the subtlety of identified causal links, or the practical utility of the generated suggestions. The qualitative assessments by domain experts, focusing on aspects like hallucination, relevance, and the quality of justifications and solutions, provided crucial insights into these aspects of performance. However, the broader field acknowledges that the robust quantification of "reasoning" in LLMs, especially for domain-specific, high-stakes tasks, is still an evolving area of research. Further investigation is imperative to develop and validate more sophisticated and reliable metrics that can truly capture the depth and accuracy of LLM-driven analytical processes.

The current investigation utilized publicly available RO-ILS cases, which provided a standardized and transparent dataset for this initial evaluation. However, the full potential of LLMs in this context likely lies in their secure, local deployment within healthcare institutions. Integrating LLMs with an institution's internal incident reporting system could enable the analysis of a much larger, more diverse, and highly contextualized dataset of local incidents. Such an approach could offer powerful insights into institution-specific vulnerabilities, help optimize local clinical workflows, identify systemic weaknesses in the chain of clinical practice, and ultimately contribute significantly to improving patient safety and the quality of care. This local application could transform RCA from a primarily reactive process to a more proactive and continuous quality improvement mechanism.

In our analysis, we observed a weak positive correlation between the length of an incident narrative, as measured by word count, and the performance metrics of our model, including recall, precision, F1-score, and accuracy (Supplementary Figure 2). Although the correlation is not strong, it suggests that more detailed descriptions can contribute to better model performances. This finding underscores the critical importance of submitting incident reports with narratives that are both accurate and comprehensive. Our observation aligns with the principles outlined by the AAPM's Task Group 288 (TG-288), which provides formal guidance for composing effective radiotherapy event narratives[49]. Adhering to guidance recommendations suggested by TG-288 can help ensure that incident learning systems receive high-quality data, thereby enhancing the potential use of LLMs for identifying vulnerabilities and improving patient safety.

Another structured safety methodology advocated by the AAPM is the TG-100, which specifically promotes Failure Mode and Effects Analysis (FMEA) as a proactive risk assessment tool in radiation therapy[50]. FMEA systematically identifies potential failure modes in treatment processes, assesses their risk based on severity, occurrence, and detectability, and prioritizes actions to mitigate these risks before incidents occur. While FMEA is highly effective, it is resource-intensive and heavily reliant on expert judgment. Our current research suggests that these LLM models could potentially support and enhance FMEA processes. Furthermore, future integration of LLM-driven approaches could assist in objectively quantifying risk factors, providing standardized, data-driven assessments that complement expert analysis, thereby potentially reducing human bias and variability.

Looking forward, there is considerable scope for enhancing the evaluation methodologies for LLM-generated RCAs. Beyond current text similarity and expert rating systems, future studies could explore the application of more advanced quantification techniques, such as those derived from causal inference frameworks. This might involve assessing the LLMs' ability to not only identify contributing factors but also to correctly map the causal pathways and interdependencies between them, perhaps even by evaluating their capacity to construct or critique causal diagrams based on incident narratives. Such methods could provide a more rigorous assessment of the depth of an LLM's understanding of the complex cause-and-effect relationships inherent in radiation oncology incidents, moving beyond correlational analysis to a more profound evaluation of genuine causal reasoning. Continued research in these directions will be vital for responsibly harnessing the power of LLMs to augment human expertise in the critical domain of patient safety.

A potential limitation of this study is the reliance on publicly available radiation oncology incident reports, which introduces the risk that these texts were included in the LLMs' training datasets (data contamination). However, the distinct variations in performance observed across the different models suggest that they are generating responses based on inference rather than merely reproducing memorized training data. To further mitigate this concern, we validated the approach using a set of internal institutional cases[51]; illustrative examples processed by Gemini 2.5 Pro are provided in Supplementary Material 5. These private cases yielded satisfactory performance consistent with the public dataset, supporting the generalizability of our findings. Nevertheless, future evaluations involving larger datasets of unseen reports remain necessary to fully confirm these results.

The reliance on public case study reports also constrains the diversity and volume of the analyzed cases. Consequently, the current dataset may not fully capture the breadth of clinical scenarios encountered across the full spectrum of radiation oncology, particularly regarding specialized modalities such as low-dose-rate (LDR) brachytherapy, adaptive radiation therapy, and proton therapy. This limitation could affect the generalizability of our findings when applied to the distinct workflows associated with these complex or less common scenarios. In future work, we plan to incorporate RO-ILS cases from our institution to enrich the dataset and expand the scope of analysis, enabling a more comprehensive evaluation of LLM performance across varied clinical contexts and treatment techniques.

## Conclusions

This study demonstrates the emerging potential of using LLMs for RCA in radiation oncology incidents. By prompting four state-of-the-art LLMs to analyze RO-ILS incident reports, we observed that these models can generate RCA outputs with high semantic similarity to expert-generated references and receive favorable evaluations from experienced medical physicists. The top-performing model achieved strong accuracy, low hallucination rates, and an overall performance rating of 4.8 out of 5, suggesting promising use in clinical applications such as workflow optimization, quality improvement and incident analysis in support of patient safety.

**Conflict of Interest Statement for All Authors**
The authors declare no conflict of interest.

**Funding Statement**
The authors declare no funding statement.

**Data Availability Statement for this Work**
Research data are stored in an institutional repository and will be shared upon request to the corresponding author.

**Acknowledgements**
This research was supported by Department of Radiation Oncology, University of Miami.

Supplementary Figure 1

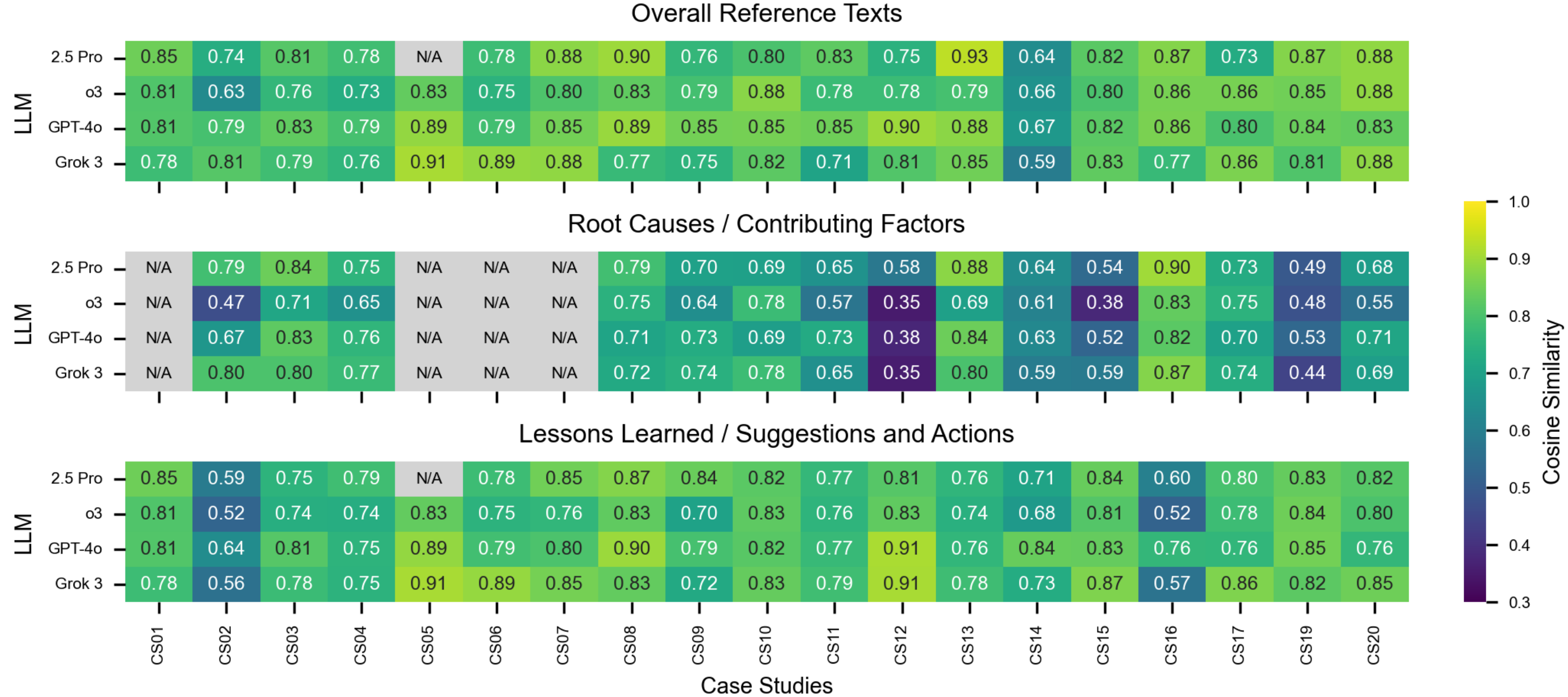

Supplementary Figure 1 Cosine similarity scores for LLM outputs compared against three sections of the reference texts: the combined sections (top), "Root Causes / Contributing Factors" (middle), and "Lessons Learned / Suggestions and Actions" (bottom).

Supplementary Figure 2

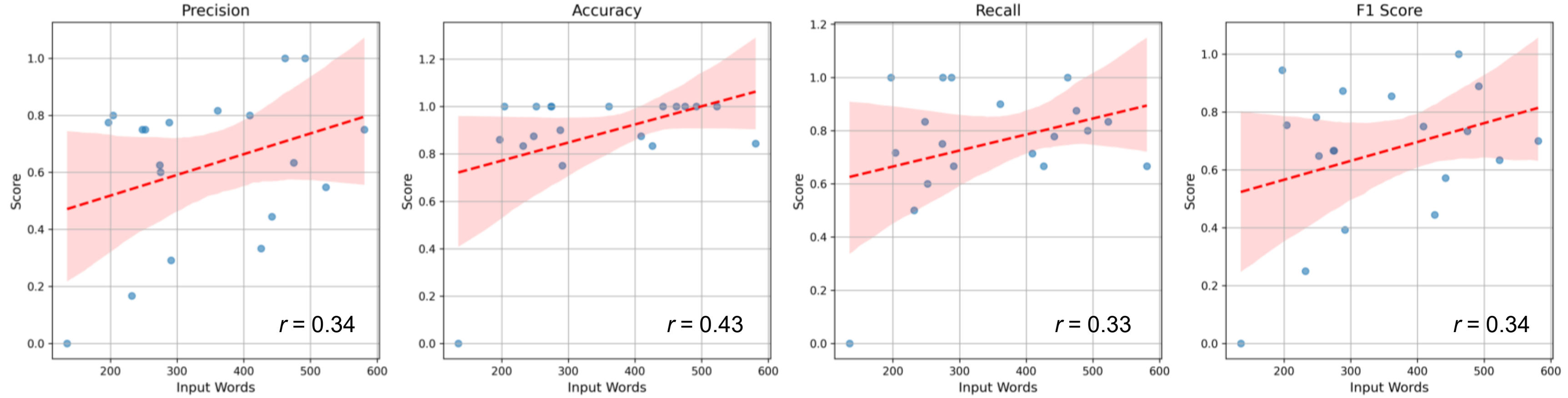

Supplementary Figure 2 Scatter plot of the prompt text length (in word count) describing radiation oncology incidents and model performance metrics, including precision, accuracy, recall, and F1-score. Pearson correlation coefficients were computed to quantify the strength of association. The red dashed line represents the linear regression fit, with the 95% confidence interval indicated by the red shaded region.

Supplemental Material 1

A representative RO-ILS case report used in this study for model analysis and evaluation

RO-ILS Case Study 02

Adaptive Planning

**Background:**

Adaptive planning (AP) has had a major impact in radiation oncology. How AP is conducted and communicated varies by institution. One common approach is patients have a quality assurance repeat CT (QACT) during the first 10 fractions and it is typically repeated at two to three times during the course of radiotherapy. The frequency and timing can vary depending upon the clinical situation and the results of the QACT. A physicist/dosimetrist performs the primary review of the QACT and this is typically done within 24 hours of obtaining the QACT. The physicist performs a rigid registration of the QACT to the original CT simulation, transfers the target and normal tissue contours to the CT simulation, and recomputes the original plan on the QACT. Dose volume histograms (DVHs) and statistics are compared between the original and re-calculated plans. An electronic document is created in the radiotherapy-specific electronic medical record and is sent to the physician to review. This QA document contains comparison DVHs, values for institutional dose constraints (target, normal tissue), and a summary recommendation from the physicist regarding the need for re-planning.

**Case Example:**

An event occurred for a patient receiving definitive radiotherapy for his intact prostate cancer. The first QACT was performed on fraction 6. The QACT document was created and sent to the physician. The physics resident reviewing the QACT did not notice that the small bowel had moved and was adjacent to the target. The max dose to the small bowel was now exceeding the institutional standard (max dose ~ 5600 cGy). The QACT document was created and sent to the physician. This document did show the comparison DVHs but the physics resident did not explicitly state the new max dose to the small bowel and did not recommend replanning. The physician also did not notice the high max dose to the small bowel. The second QACT was performed on fraction 22. The same loop of small bowel that had moved adjacent to the target was again in the same location and the max small bowel dose was again exceeding institutional standards (max ~5600 cGy) but was not noticed by the physics resident. The second QACT document also did not explicitly highlight the new max dose to the small bowel. The physician reviewed the second QA document and this time noticed the high small bowel dose and requested a repeat QACT to occur as soon as possible. This request was communicated via adding a comment to the second QACT document and sending it back to the physics resident. The physics resident did not notice this request from the physician. At the time for planning the boost phase of treatment, the dosimetrist reviewed the QACT and noted the high max dose to the small bowel. Subsequently, the boost plan was optimized using a tighter objective goal for the small bowel, accounting for the high dose the small bowel received during the initial course.

**Contributing Factors/Root Causes:**

1. Workload: Adaptive planning is done routinely on all patients at some institutions which significantly increases the workload. Most QACTs require no action (i.e., re-planning). Suboptimal workload (in this case too high) may have contributed to the physics resident and physician inadvertently not noticing the high dose to small bowel [1, 2].
2. Documentation: The QACT is an automated document and there was no field for reporting the max dose to small bowel. DVHs can be difficult to review, especially when there are many structure sets overlaid in one graph.
3. Communication: There was no standard method for physicians to communicate the order for repeat QACT.
4. Process: For the initial CT simulation the individual loops of small bowel were contoured, but for the subsequent QACTs a bowel space was contoured. If the patient's small bowel loops were contoured on the QACT then it may have improved the acuity of the physicist to notice that a loop of the small bowel had moved to be adjacent to the target.
5. Supervision: A physics resident was conducting the review of the QACTs. A second check by a supervising physicist may have prevented this event.

**Lessons Learned:**

Adaptive planning is quickly becoming a standard part of the radiation oncology process (e.g., head and neck IMRT, MRI LINAC, proton therapy). The obvious value is in improving the precision/accuracy of radiation treatments, thus potentially reducing toxicity/morbidity and improving cancer control. However, there is a significant cost in that adaptive planning increases workload. Standard process and procedures as well as the automation of treatment planning and QA checks must be developed to improve the effectiveness and efficiency of adaptive planning. This event illustrates the need for the design of robust processes for the effective and efficient implementation and communication of adaptive planning.

Supplemental Material 2

Evaluation table and rubric used by medical physicists for assessing LLM-generated outputs

| | | | Model 1 | Model 2 | Model 3 | Model 4 |
|---|---|---|---|---|---|---|
| SEMI_SUBJECTIVE | | | | | | |
| Accuracy and Correctness | (Focus on Root Cause and Contributing Factors) | | | | | |
| | **Precision:** Of the items the LLM identified, how many were mentioned in CaseStudy? | # of LLM identified items in Case Study | | | | |
| | | # of LLM identified items are correct | | | | |
| | | # of LLM identified items | | | | |
| | | | #DIV/0! | #DIV/0! | #DIV/0! | #DIV/0! |
| | **Recall:** Of all the items identified in CaseStudy document, how many did the LLM identify? | # of LLM identified items in CaseStudy | 0 | 0 | 0 | 0 |
| | | # of total items identified in CaseStudy | | 0 | 0 | 0 |
| | | | #DIV/0! | #DIV/0! | #DIV/0! | #DIV/0! |
| | **F1 Score:** Harmonic mean of precision and recall. | | #DIV/0! | #DIV/0! | #DIV/0! | #DIV/0! |
| Faithfulness and Hallucination | (Focus on whole answers) | | | | | |
| | **Hallucination:** Identifies if the LLM's response includes fabricated or inaccurate information. | 1-Yes (Fabricated info identified); 0-No (No fabricated info identified) | | | | |
| Relevance and Comprehensiveness | (Focus on the whole answers) | | | | | |
| | **Relevance:** Evaluates whether the model's responses are pertinent to the specific case (not a general answer) | 1-Fully Case Unrelavant (Relavance<5%); 2-Mostly Case Unrelavant (Relavance<25%); 3-Somehow Case Relavant (Relavance about 50%); 4-Mostly Case Relavant (Relavance>75%); 5-Fully Case Relavant (Relavance>95%) | | | | |
| | **Comprehensiveness:** Evaluate whether the identified root causes and actions are comprehensive | 1-Very Poor (Miss almost all points); 2-Poor (Get some points); 3-Neutral (Get about half of points); 4-Good (Get most points); 5-Excellent (Get almost all the points I can think) | | | | |
| Justifications and Solutions | (Focus on the analysis and suggestion/action) | | | | | |
| | **Quality of Justification:** Assesses the model's ability to provide clear, logical, and coherent explanations for the identified root causes | 1-Very Poor (Justification is very unclear, illogical and incoherent); 2-Poor (Justification is not clear, logical and coherent); 3-Neutral (Justification can barely address the point); 4-Good (Justification is mostly clear, logical and coherent); 5-Excellent (Justification is very clear, logical and coherent) | | | | |
| | **Quality of Solutions:** Assesses the model's ability to provide reasonable, actionable, and relevant suggestions and actions | 1-Very Poor (Almost all of the suggestions are unreasonable); 2-Poor (Most of suggestions are unreasonable); 3-Neutral (Half of the suggestions are reasonable); 4-Good (Most of the suggestions are reasonable); 5-Excellent (Almost all of the suggestions are reasonable) | | | | |
| SUBJECTIVE | | | | | | |
| Human Metrics | | | | | | |
| | **Logical Reasoning Score** | 1-Very Poor (Almost no logical reasoning); 2-Poor (Poor performance of logical reasoning); 3-Neutral (Show logical reasoning in some parts); 4-Good (Good performance of logical reasoning); 5-Excellent (Comparable to human in logical reasoning) | | | | |
| | **Overall Performance Score** | 1-Very Unsatisfied; 2-unsatisfied; 3-Neutral; 4-Satisfied; 5-Very Satisfied | | | | |

Supplemental Material 3

Representative output samples generated by the four evaluated LLMs

1. Gemini Pro 2.5 (Google)

**Root Cause Analysis (RCA) of Adaptive Planning Event**

**Event Summary:** During a course of definitive radiotherapy for prostate cancer, routine Quality Assurance CT (QACT) scans were performed for adaptive planning. On two separate occasions (fraction 6 and fraction 22), a physics resident reviewing the QACTs failed to identify and explicitly report a significant increase in the maximum dose to the small bowel, which had moved adjacent to the target and exceeded institutional dose constraints. The initial QACT documentation, while including comparison DVHs, did not highlight this critical finding or recommend re-planning. The physician also missed the high small bowel dose on the first QACT review. Upon reviewing the second QACT, the physician noticed the high dose and requested a repeat QACT via a comment in the electronic document sent back to the physics resident; however, the resident missed this request. The high small bowel dose was ultimately identified by a dosimetrist during boost planning, and the boost plan was subsequently optimized to compensate.

---

**1. Chronological Sequence of Events**

This section diagrams the flow of events leading up to the incident, incorporating the "three whys" (What happened? Why did it happen? What are you going to do about it?) at critical junctures.

1. **Initial Condition:** Patient undergoing definitive radiotherapy for intact prostate cancer, with a standard institutional adaptive planning (AP) process involving QACTs.
    - **What happened?** Standard AP process initiated.
    - **Why did it happen?** Part of the institution's quality assurance and plan adaptation strategy to account for anatomical changes during treatment.
    - **What are you going to do about it?** Ensure the AP process has robust checks and communication pathways.
2. **Event 1: First QACT Review (Fraction 6)**
    - Patient has first QACT on fraction 6.
    - Physics resident performs primary review: rigid registration, contour transfer, re-computation of original plan on QACT.
    - Small bowel has moved adjacent to the target, resulting in the max dose to the small bowel exceeding institutional standard (max dose ~ 5600 cGy).

- The physics resident *does not notice* the small bowel has moved or that the max dose is exceeding the standard.
- QACT document created with comparison DVHs but *without explicit statement* of the new max dose to small bowel and *without a recommendation for re-planning*.
- Document sent to the physician.
- The physician reviews the QACT document but also *does not notice* the high max dose to the small bowel.
- **What happened?** Critical information (small bowel overdose) was missed by both the physics resident and the physician during the first QACT review.
- **Why did it happen? (1st Why - Physics Resident)** Possible inexperience, lack of focused checklist for critical OARs, cognitive error (e.g., search satisfaction after checking target coverage), or unclear presentation of critical data in the QACT document template.
- **Why did it happen? (1st Why - Physician)** Possible time constraints, reliance on physics summary, QACT document format not clearly highlighting critical OAR violations, or cognitive error.
- **Why did it happen? (2nd Why - System/Process for both)** Lack of a structured, mandatory checklist for QACT review that specifically requires verification and documentation of all critical OAR constraints. QACT document format may not optimally highlight out-of-tolerance values. Insufficient emphasis or training on specific potential anatomical changes known to occur (e.g., small bowel mobility in prostate treatment).
- **Why did it happen? (3rd Why - Organizational/Cultural for both)** Potential gaps in supervision or direct oversight for residents performing critical QACT reviews. Culture may not sufficiently emphasize independent double-checks for critical QACT parameters. Lack of clear accountability for identifying and flagging constraint violations.
- **What are you going to do about it?** Implement standardized QACT review checklists, improve QACT report formatting to highlight deviations, enhance training on OAR changes, review supervision protocols for residents, and implement clear alert mechanisms for constraint violations.

3. **Event 2: Second QACT Review (Fraction 22)**

- Patient has second QACT on fraction 22.
- Physics resident performs review (same process as Event 1).
- The same loop of small bowel is in the same problematic location, and the max small bowel dose again exceeds institutional standards (~5600 cGy).
- The physics resident again *does not notice* the high small bowel dose.
- Second QACT document created and sent to the physician, again *without explicit highlighting* of the new max dose to the small bowel.
- **What happened?** The same critical information (small bowel overdose) was missed again by the physics resident.
- **Why did it happen? (1st Why)** Similar reasons as in Event 1 (inexperience, lack of checklist, cognitive error). Failure to learn from or identify patterns from the previous QACT (though it wasn't recognized then either).
- **Why did it happen? (2nd Why - System/Process)** Persistence of the same systemic weaknesses identified in Event 1. Lack of feedback loop or longitudinal review of QACTs that might flag persistent issues.
- **Why did it happen? (3rd Why - Organizational/Cultural)** Persistence of the same organizational/cultural factors identified in Event 1.
- **What are you going to do about it?** Reinforce actions from Event 1. Consider systems for longitudinal tracking of QACT metrics.

4. **Event 3: Physician Identifies Issue on Second QACT and Attempts Communication**
    - The physician reviews the second QACT document.
    - This time, the physician *notices* the high small bowel dose.
    - Physician requests a repeat QACT "as soon as possible" by *adding a comment to the second QACT document and sending it back to the physics resident* within the electronic medical record system.
    - The physics resident *does not notice this request* from the physician.
    - **What happened?** The physician identified the problem but the chosen communication method for an urgent request failed.

- **Why did it happen? (1st Why)** The electronic communication method (comment in a returned document) was not sufficiently robust or actively monitored for timely action by the resident. The physician may have assumed this method was adequate for an ASAP request.
- **Why did it happen? (2nd Why - System/Process)** Lack of a standardized, acknowledged, and tracked critical findings communication protocol. The system may not have a clear notification mechanism for comments added to returned documents, or the resident was not trained/expected to monitor for such feedback actively. No closed-loop communication was ensured for an "ASAP" request.
- **Why did it happen? (3rd Why - Organizational/Cultural)** Over-reliance on passive electronic communication for potentially urgent clinical findings. Lack of clear institutional guidelines on preferred communication methods for escalating concerns or requesting urgent actions based on QACT reviews.
- **What are you going to do about it?** Implement a standardized, direct, and acknowledged communication protocol for critical/urgent findings from QACT reviews. Improve EMR notification systems or establish alternative direct communication pathways for such issues.

5. **Event 4: Issue Identified During Boost Planning**
    - At the time of planning the boost phase of treatment, the dosimetrist reviews the previous QACTs.
    - The dosimetrist *notes* the high max dose to the small bowel from the initial course QACTs.
    - The boost plan is subsequently optimized using a tighter objective goal for the small bowel to account for the high dose received during the initial course.
    - **What happened?** The problem was finally identified and partially mitigated at a later stage by a different team member.
    - **Why did it happen?** The dosimetrist's routine review process for boost planning included a comprehensive look at prior imaging and dosimetry, providing another safety net.
    - **What are you going to do about it?** While good that it was caught, this is a late catch. Actions should focus on earlier detection. However, acknowledge

the value of this existing safety check in the boost planning phase and ensure its robustness.

---

## 2. Cause and Effect Diagramming

Identifying the conditions that resulted in the failure to recognize and act upon the high small bowel dose in a timely manner.

**Categories of Potential Causes:**

- **People:**
    - *Physics Resident:*
        - Failed to detect small bowel proximity/overdose in QACT 1 & 2 (Possible inexperience, insufficient attention to detail, lack of specific focus on OAR changes, knowledge gap, confirmation bias focused on target).
        - Failed to explicitly document the (unrecognized) high dose.
        - Failed to recommend re-planning (as the issue was unrecognized).
        - Missed physician's electronic request for repeat QACT (communication system failure, not actively checking for feedback).
    - *Attending Physician:*
        - Failed to detect small bowel overdose in QACT 1 (Possible over-reliance on physics summary, time constraints, report format, cognitive error).
        - Used a passive communication method for an urgent request (comment in returned document).
    - *Dosimetrist:*
        - Successfully identified the issue at a later stage (acting as a safety net).
- **Processes/Procedures:**
    - *QACT Review Process:*

- Lack of a standardized, detailed checklist for physics review of QACTs, especially for trainees (e.g., mandating check of all OARs against constraints, explicit note of organs moving into high dose regions).
        - Inadequate procedure for escalating or highlighting critical OAR dose violations.
        - No clear process for independent second check of resident's QACT review by senior physicist/dosimetrist, especially for initial QACTs.
    - *QACT Documentation/Reporting:*
        - Standard QACT document format may not effectively highlight critical OAR violations (e.g., no color-coding or automated flagging of exceeded constraints).
        - Expectation for narrative summary from physicist might be insufficiently defined regarding what to explicitly state.
    - *Communication Protocol:*
        - Lack of a robust, standardized, and acknowledged (closed-loop) communication protocol for critical or urgent findings from QACT reviews (e.g., physician's request for repeat QACT). Reliance on passive EMR comments for urgent matters.
        - No clear expectation for how quickly physics should review physician feedback on QACT documents.
    - *Supervision & Training:*
        - Potentially insufficient direct oversight or structured feedback mechanism for physics residents performing QACT reviews.
        - Training may not sufficiently emphasize common anatomical changes and their dosimetric impact in specific disease sites (e.g., small bowel in prostate RT).
- **Technology/Equipment (EMR/Reporting Tools):**
    - *EMR System:*
        - May lack effective notification mechanisms for comments added to returned/updated documents.

- QACT report template within the EMR may not be optimized for clarity of critical data (e.g., visual cues for constraint violations).
        - Lack of automated flagging of OAR constraint violations within the re-calculated plan display or DVH comparison.
    - *Treatment Planning System (TPS) used for QACT re-computation:* While it provided DVHs, it did not prevent the oversight directly without active user interpretation.
- **Organizational/Environmental Factors:**
    - *Safety Culture:*
        - Potential for diffusion of responsibility (e.g., physician assuming physics fully vetted all OARs, physics assuming physician will catch issues on DVHs).
        - Lack of a strong culture of proactively communicating and confirming receipt of critical/urgent information.
    - *Workload/Time Pressures:* (Potential, not explicitly stated) May affect thoroughness of review for all staff involved.
    - *Clarity of Roles & Responsibilities:* Ambiguity in who is ultimately responsible for explicitly noting *every* OAR violation versus providing a general summary.
    - *Resident Supervision Model:* May need review to ensure adequate checks and balances for tasks with direct patient safety implications.

---

**3. Causal Statements, Actions, and Outcomes**

**Root Cause(s):**

1. **RC1: Inadequate Standardized QACT Review and Reporting Process, Lacking Mandatory Checks and Clear Highlighting of Critical Deviations.** The process for QACT review and documentation did not ensure consistent identification and prominent communication of critical organ-at-risk (OAR) dose constraint violations. This was evidenced by the physics resident repeatedly missing the small bowel overdose and the QACT document failing to explicitly flag this critical finding for the physician.

- **Contributing Factor:** Lack of a mandatory, detailed checklist for QACT review by physics staff (especially residents) that requires explicit verification and documentation of all critical OAR constraints and anatomical changes.
- **Contributing Factor:** QACT report format not optimized to automatically and visually flag or highlight OAR dose violations, making them easier to overlook in a complex data presentation.
- **Contributing Factor:** Insufficient training or emphasis on common site-specific anatomical variations (e.g., small bowel migration in prostate treatment) and their dosimetric consequences during QACT review.

2. **RC2: Deficient Communication Protocol for Critical/Urgent QACT Findings and Follow-up Actions.** The communication pathway for reporting and responding to critical QACT findings, such as the physician's request for an urgent repeat QACT, was not robust, timely, or actively monitored, leading to a missed request.
    - **Contributing Factor:** Reliance on passive electronic communication (e.g., comments in a returned document within the EMR) for urgent clinical requests without a clear notification or mandatory acknowledgment loop.
    - **Contributing Factor:** Lack of a clearly defined institutional protocol for escalating urgent requests stemming from QACT reviews, specifying preferred communication methods and expected response times.

**Contributing Factor(s) (not direct root causes but enabling the event):**

- **CF1: Inadequate Supervision and/or Verification of Physics Resident QACT Reviews:** The physics resident, potentially due to inexperience, missed critical findings on two occasions. The system lacked a sufficiently robust checkpoint or secondary review for resident-performed QACTs before the information (or lack thereof) reached the physician.
- **CF2: Physician Oversight/Miss in First QACT Review:** The physician also initially missed the high small bowel dose on the first QACT, indicating a potential over-reliance on the physics summary or a gap in their own detailed review of the provided DVHs/images at that time.

**Actions and Outcome Measures:**

1. **Action (Addressing RC1):** Develop and implement a standardized, mandatory QACT Review Checklist and Reporting Template.
    - **Specifics:**

- Checklist to include: explicit verification of all institutional OAR dose constraints, notation of significant anatomical changes (e.g., OAR proximity to target), and required physicist comments on any deviations.
        - New reporting template to automatically highlight or use distinct visual cues (e.g., color-coding, bold text) for any OAR values exceeding constraints in the QACT document.
        - The physicist's summary recommendation must explicitly state if re-planning is or is not recommended and *why*, referencing specific OARs if they are the driver.
    - **Outcome Measure:** 100% utilization of the new checklist and template for all QACT reviews. Audit of QACT documents for clarity and completeness. Reduction in missed OAR violations to zero in subsequent QACT reviews.
2. **Action (Addressing RC1 & CF1):** Enhance Training and Supervision for QACT Review.
    - **Specifics:**
        - Conduct specific training for all physics staff (especially residents) on common site-specific anatomical variations and their dosimetric impact, using case examples like this one.
        - Implement a policy requiring a countersignature or documented secondary review by a senior physicist/dosimetrist for all QACTs reviewed by residents, particularly for the initial QACTs in a patient's course or if significant changes are noted.
    - **Outcome Measure:** Documentation of completed training and adherence to secondary review policy. Improved accuracy in resident QACT assessments.
3. **Action (Addressing RC2):** Establish a Standardized "Critical QACT Findings Communication Protocol."
    - **Specifics:**
        - Define what constitutes a critical/urgent QACT finding (e.g., OAR exceeding x% of constraint, physician request for immediate action).
        - Mandate direct, person-to-person communication (e.g., phone call, secure message requiring acknowledgment) for such findings, in

- addition to EMR documentation. Document who was contacted, when, and the agreed-upon action.
        - Discourage reliance on passive EMR comments for urgent requests. If EMR is used for non-urgent feedback, ensure robust notification systems are active and staff are trained to monitor them.
    - **Outcome Measure:** 100% adherence to the new communication protocol for critical findings. Audit trail of acknowledged communications. No missed urgent requests.
4. **Action (Addressing CF2 & Physician Review):** Physician QACT Review Enhancement.
    - **Specifics:** Discuss this event with physicians to reinforce the importance of their independent detailed review of QACT data (DVHs, images), even when a physics summary is provided. Consider developing a physician-specific QACT review checklist or guidance points.
    - **Outcome Measure:** Increased physician awareness and engagement in detailed QACT review, potentially measured via feedback or a reduction in reliance solely on physics summaries for critical OARs.
5. **Action (Technology Enhancement):** Explore EMR/TPS Enhancements for QACT.
    - **Specifics:** Work with EMR/TPS vendors or internal IT to:
        - Improve automated notifications for document updates/comments requiring action.
        - Develop capabilities for automated flagging of OAR constraint violations directly within the plan review or DVH comparison tools used for QACT.
    - **Outcome Measure:** Implementation of any feasible technological improvements within a defined timeframe (e.g., 12-18 months).

---

**Lessons Learned from this Incident**

- **Adaptive Planning Complexity:** AP introduces new decision points and data reviews; if processes are not robust, errors of omission (missing critical changes) can occur.

- **Silent Failures:** The failure to detect the small bowel overdose was a "silent" error that persisted through multiple checks until a later stage, highlighting the danger of process gaps.
- **Trainee Supervision:** Critical tasks performed by trainees, such as QACT reviews, require structured oversight and verification to ensure patient safety, especially when findings directly impact treatment decisions.
- **Data Overload vs. Key Information:** QACT reports can contain extensive data (DVHs, statistics). Critical information (like an OAR overdose) must be explicitly and clearly highlighted, not just present amidst other data.
- **Communication is More Than Sending:** Effective communication requires ensuring the message is received, understood, and, if urgent, acknowledged with a plan of action. Passive electronic comments are often insufficient for critical/urgent matters.
- **Diffusion of Responsibility:** When multiple professionals review data, there's a risk each might assume others have caught or will catch an issue. Clear accountability at each step is vital.
- **Importance of Multiple Safety Nets:** The dosimetrist catching the issue during boost planning demonstrates the value of built-in redundant checks at different points in the radiotherapy workflow. However, reliance on late-stage catches is not ideal.
- **EMR as a Tool, Not a Panacea:** EMRs facilitate information sharing, but their design for notifications and highlighting critical data can significantly impact workflow safety. User practices must adapt to system limitations.

---

**Suggestions and Actions**

- **Standardize QACT Review:** Implement a mandatory, detailed checklist for all physics staff performing QACT reviews, requiring explicit verification of target coverage and *all* critical OAR constraints against institutional limits.
- **Optimize QACT Reporting:** Revise QACT documentation templates to:
    - Automatically flag or use prominent visual cues (e.g., red/yellow highlighting) for any parameters exceeding thresholds (e.g., OAR doses, target underdosing).

- Require a clear, concise narrative summary that explicitly states whether re-planning is indicated and details any significant OAR changes or dose violations.

- **Implement Tiered Review for QACTs:** For QACTs reviewed by residents or junior staff, institute a mandatory secondary review and sign-off by a senior physicist or dosimetrist, especially for initial QACTs or those showing significant deviations.
- **Establish Clear Communication Protocols for Urgent Findings:**
    - Define what constitutes an urgent/critical QACT finding.
    - Mandate direct, acknowledged communication (e.g., phone call, direct secure message with read-receipt) for such findings to the responsible physician, supplementing EMR documentation.
    - Set clear expectations for response times to urgent QACT-related requests.
- **Enhance EMR Notifications:** Work with IT/vendors to improve EMR systems to provide more active and obvious notifications for clinically relevant updates, such as comments on shared documents or critical results.
- **Regular Interdisciplinary Case Reviews:** Periodically review adaptive planning cases (especially those requiring intervention or where issues were noted) in interdisciplinary meetings (physics, dosimetry, physicians) to share learning and identify process improvement opportunities.
- **Specific Training on Anatomical Changes:** Provide focused training on common and critical anatomical variations observed during radiotherapy for different disease sites (e.g., bowel gas/filling changes in pelvic RT, tumor shrinkage in H&N RT) and their dosimetric implications.
- **Physician QACT Review Aids:** Consider developing a brief checklist or key "look-for" items to aid physicians in their review of QACT documentation, ensuring they cross-verify critical OARs even when a physics summary is provided.
- **Promote a "Questioning Attitude":** Encourage all team members, regardless of experience, to voice concerns or seek clarification if QACT results seem unclear, unexpected, or if documentation lacks explicit statements on critical parameters.
- **Longitudinal QACT Data Review:** Where feasible, implement methods to easily review QACT data longitudinally for a patient, which might help identify trends or persistent issues that were subtle on a single QACT.

2. o3 (OpenAI)

## Root-Cause Analysis (RCA) – Adaptive-Planning Missed Small-Bowel Over-Dose

*Structured per the AAPM "Simple Framework for RCA"*

---

### 1 Chronological sequence (event flow & the "three whys")

| Step | Date / fx | What happened | Why #1 (immediate) | Why #2 (underlying) | Why #3 (root) |
|---|---|---|---|---|---|
| 1 | fx 6 | QACT-1 acquired; physics resident creates QA report | Resident **misses high max dose** (≈ 56 Gy) to small bowel | DVH table present, but no *forced highlighting* of constraint violations | QA template lacks automated red-flag / hard-stop for constraint breaches |
| 2 | same day | QA report e-signed & routed to physician | Resident's comment **omits explicit mention** of bowel violation | No checklist item "State *all* constraint deviations" | Training & SOP do not require narrative of every failed metric |
| 3 | fx 6+1 | Physician reviews report, **also misses** violation | DVH graphic crowded; bowels not contoured in colour key | Template design & reading workflow overload visual search | Human-factor design of report not usability-tested |
| 4 | fx 22 | QACT-2 acquired; resident again completes QA | Same loop of small bowel still high (> 56 Gy); **missed again** | Heavy workload; resident assumes "no change" if plan unchanged | No second-checker or peer review of QACTs |
| 5 | same day | Physician reviews second QA; **spots violation** | Adds comment: "Repeat QACT ASAP" & returns doc to physics | Communication embedded as comment; no alert to resident | EMR lacks task/alert system for returned QACT documents |
| 6 | fx 22+1 | Resident **does not see** | No dashboard/list for "documents needing action" | Resident depends on email pop-ups, | Electronic workflow governance |

| Step | Date / fx | What happened | Why #1 (immediate) | Why #2 (underlying) | Why #3 (root) |
|---|---|---|---|---|---|
| | | physician’s request | | which failed due to filter | missing; reliance on email only |
| 7 | Boost planning | Dosimetrist reviews QACTs; notes violation | Escalates → boost plan re-optimised with tight bowel objective | Discovery late but before further harm | Defensive dose objective compensates but cannot undo initial dose |

---

**2 Cause-and-Effect diagram (textual Ishikawa)**

pgsql

CopyEdit

```
                PATIENT RISK: Small-bowel dose > institutional max

                    ______________________________________

  PEOPLE         PROCESS        TECHNOLOGY       ENVIRONMENT     MEASUREMENT

 • Single-review  • No forced     • QACT template   • High workload  • DVH limits not

  resident        two-person      does not flag     on residents    auto-checked

  reviewer        check          violations       • Alert via EMR  • No report metric

 • Physician      • No checklist  • EMR comments      comments        dashboard

  relies on       item “state     lack push alert              • No audit trail

  physics         all failures”  • No dashboard

                    ______________________________________
```

---

**3 Causal statements**

| ID | Causal statement (Cause → Effect → Rationale) | Type |
|---|---|---|
| CS-1 | *If* the QACT QA template **does not automatically flag or block constraint violations**, *then* reviewers may overlook high doses, *because* the deviation is buried in routine DVH plots. | **Root cause** |
| CS-2 | *If* only one qualified person reviews each QACT, *then* detection of anatomy changes is dependent on one individual's vigilance, *because* a mandated second-check is absent. | Root cause |
| CS-3 | *If* communication from physicians relies on **embedded EMR comments without active alerts**, *then* action requests can go unnoticed, *because* the system does not generate tasks or notifications. | Contributing |
| CS-4 | *If* the QACT checklist **does not require explicit narrative** of every failed constraint, *then* reviewers may omit critical information, *because* the template suggests optional commentary. | Contributing |
| CS-5 | *If* resident workload is high and no time is allotted for careful adaptive review, *then* cognitive oversights increase, *because* of time pressure and fatigue. | Contributing |

---

**4 Lessons learned**

- Automated, **hard-stop dose-constraint checks** are more reliable than manual DVH inspection.
- A **two-person or physics-MD peer-review** model materially lowers the miss rate for adaptive QA.
- Free-text comments are insufficient; EMRs need **task/alert mechanisms** tied to adaptive-planning workflows.
- Checklists must force reviewers to **state every metric out-of-tolerance** (pass/fail logic).
- Human-factor design (colour-coded violation banners) improves rapid recognition over raw DVH plots.
- Training should emphasise that **bowel position can change early** in prostate RT, warranting vigilance.

- Workload management and protected time are critical for careful adaptive reviews.

---

**5 Suggestions & actions**

| # | Action | Owner | Deadline | Success metric |
| --- | --- | --- | --- | --- |
| 1 | Integrate **auto-validation script** in TPS/EMR: any dose ≥ constraint turns field red & blocks sign-off until acknowledged. | Physics IT | 45 days | Script live; 100 % of future deviations auto-flagged |
| 2 | Implement **mandatory second-check** (peer physicist or dosimetrist) for every QACT within 24 h. | Chief Physicist | 14 days | Audit: ≥ 95 % QACTs have documented second checker |
| 3 | Revise QACT template: add **pass/fail table**; require comment for any "fail." | QA Committee | 30 days | New template in use; staff survey shows clarity ↑ |
| 4 | Configure EMR to create a **task with pop-up alert** when a document is returned for action. | IT / Vendor | 60 days | Alert log shows 0 missed tasks after implementation |
| 5 | Provide **focused training** for residents on GI-organ motion and adaptive triggers. | Residency Director | Next cycle | Post-test score ≥ 90 %, competencies logged |
| 6 | Quarterly **workload review** of adaptive QA duties; adjust staffing or scheduling as needed. | Department Admin | Ongoing | Turn-around time ≤ 24 h with <10 % overtime |

---

These corrective actions couple **system improvements** (automation, checklists, alerts) with **human-factor fixes** (peer review, training, workload) to prevent recurrence and strengthen the adaptive-planning safety net.

3. GPT-4o (OpenAI)

Here is a **standardized Root Cause Analysis (RCA)** for **RO-ILS Case Study 02: Adaptive Planning**, structured according to the **AAPM** framework:

## 1. Chronological Sequence (Including the Three "Whys")

### Timeline of Events

```
text
CopyEdit
Fraction 6:
↓
QACT performed → Physics resident reviews → Small bowel moved adjacent to
target
↓
Dose recalculated (max ~5600 cGy to small bowel) → Not highlighted in the QA
document
↓
Physician reviews QACT but misses small bowel dose → Treatment continues

Fraction 22:
↓
Second QACT performed → Same small bowel location and dose (~5600 cGy)
↓
Physics resident again misses elevated small bowel dose → QA document again
lacks emphasis
↓
Physician catches dose on second review → Requests repeat QACT (via document
comment)
↓
Physics resident misses the physician's comment
↓
At boost planning, dosimetrist notices small bowel dose issue
↓
Boost plan adapted to reduce further small bowel dose
```

### The Three "Whys"

- **Why did the small bowel receive a dose exceeding institutional constraints?**
  → Because its new position adjacent to the target in QACTs was not identified as problematic and no adaptive replan was initiated.

- **Why was adaptive replanning not initiated?**
  → Because the physics resident reviewing the QACTs did not highlight the elevated small bowel dose, and the physician missed it in the first review.
- **Why did the communication breakdown occur regarding replan requests and dose thresholds?**
  → Because the QA documentation lacked clear escalation or alerts for constraint violations, and passive communication (via comments) was ineffective.

# 🔍 2. Cause and Effect Diagramming (Ishikawa/Fishbone Analysis)

## Main Categories and Contributing Factors

### *People (Human Factors)*

- Inexperienced physics resident missed clinically significant findings.
- Physician did not catch elevated dose on initial review.
- Missed communication regarding replan request due to reliance on passive messaging.

### *Process*

- No standardized checklist or flagging system for dose constraint violations in QACT reviews.
- No formal communication protocol for adaptive planning decisions or physician replan requests.

### *Technology*

- QA documents not designed to clearly highlight dose constraint violations.
- No alert system to notify when a comment or request is made in a document.

### *Environment*

- Potential workload or time pressure on physics and physician teams.
- Non-standardized training or expectations for QACT review responsibility and rigor.

# 3. Causal Statements

## Root Cause Statement

- **The failure to identify and communicate an adaptive planning need resulted from unclear processes for flagging dose constraint violations in QACT review documents and reliance on informal communication methods.**

## Contributing Factors

- Lack of formal review checklist for identifying dose violations.
- Inadequate training or supervision of junior staff in QACT review.
- Passive communication (document comments) insufficient for clinical decision-making.
- Absence of an alert system when DVH metrics exceed institutional constraints.

## Outcome

- The error was caught before further dose delivery due to the dosimetrist's diligence during boost planning.
- Patient received an optimized boost plan, minimizing further risk.
- No acute clinical toxicity was reported, but this incident represented a near miss with potential for significant harm.

# Lessons Learned (Bullet Points)

- Junior staff reviewing QACTs require structured oversight and training in identifying critical organ-at-risk (OAR) dose violations.
- Passive communication (e.g., comments in documents) is not reliable for time-sensitive or high-risk clinical decisions.
- QA documents should clearly flag any violation of institutional dose constraints.
- Clinician oversight must be vigilant, even if initial physics review does not recommend replanning.
- Multidisciplinary communication is essential for adaptive radiotherapy, especially when anatomy changes.
- Repetition of the same error on subsequent QACTs suggests process failure, not individual oversight alone.
- QACT review is not a routine task but a high-stakes quality assurance process that demands formalization.

## ☑ Suggestions and Actions (Recommendations)

- **Implement a mandatory checklist** for QACT reviews that requires explicit documentation of all OAR doses vs. institutional constraints.
- **Design QA documents to automatically flag constraint violations** in red or with an alert icon for immediate visibility.
- **Establish a formal escalation protocol**: any plan exceeding constraints must be escalated to attending physician and chief physicist within 24 hours.
- **Train and supervise physics residents** more closely in adaptive planning procedures and DVH analysis.
- **Replace comment-only communication** with direct messages, task assignments, or alerts in the EMR or planning system.
- **Audit QACT reviews periodically** to assess accuracy, communication clarity, and outcomes.
- **Encourage a multidisciplinary adaptive planning huddle** (physician, physicist, dosimetrist) for complex or evolving cases.
- **Integrate automated systems** that compare recalculated dose with constraints and generate alerts when exceeded.

Let me know if you'd like this RCA output adapted into a PowerPoint or PDF format for teaching or documentation purposes.

4. Grok 3 (Grok)

# Root Cause Analysis: Adaptive Planning Error in Radiation Oncology

## Chronological Sequence

The following outlines the sequence of events leading to the incident where a high dose to the small bowel was not addressed promptly during adaptive planning for a patient receiving radiotherapy for intact prostate cancer:

1. **Initial Treatment Planning**: The patient began definitive radiotherapy for intact prostate cancer, with a treatment plan including quality assurance CT (QACT) scans scheduled during treatment to monitor anatomical changes.
2. **First QACT (Fraction 6)**: A QACT was performed on fraction 6. The physics resident reviewed the QACT, performed a rigid registration to the original CT simulation, transferred contours, and recomputed the original plan. The small bowel had moved adjacent to the target, resulting in a maximum dose exceeding the institutional standard (~5600 cGy). The physics resident did not notice this, and the QACT document, while including comparison dose-volume histograms (DVHs), did not explicitly highlight the high small bowel dose or recommend replanning.
3. **Physician Review of First QACT**: The physician reviewed the QACT document but did not notice the high small bowel dose, and treatment continued without replanning.
4. **Second QACT (Fraction 22)**: A second QACT was performed on fraction 22. The same small bowel loop remained adjacent to the target, with the max dose again exceeding institutional standards (~5600 cGy). The physics resident again failed to note the high dose, and the QACT document did not explicitly highlight it or recommend replanning.
5. **Physician Review of Second QACT**: The physician reviewed the second QACT document, noticed the high small bowel dose, and requested an immediate repeat QACT by adding a comment to the document and sending it back to the physics resident.
6. **Missed Communication**: The physics resident did not notice the physician's request for a repeat QACT, and no immediate action was taken.
7. **Boost Phase Planning**: During planning for the boost phase, the dosimetrist reviewed the QACT documents, identified the high small bowel dose, and optimized the boost plan with a tighter objective goal for the small bowel to account for the excess dose received earlier.

**Three "Whys" Analysis**:

- **Why was the high small bowel dose not addressed promptly?** The physics resident did not identify the high small bowel dose in the QACT reviews, and the QACT documents did not explicitly highlight the issue, delaying physician awareness until fraction 22.

- **Why did the physics resident miss the high dose and fail to highlight it?** There was no standardized checklist or protocol requiring explicit documentation of dose constraint violations in QACT reports, and the resident may have lacked sufficient training or oversight.
- **Why was there no standardized protocol or adequate training?** The adaptive planning process lacked clear guidelines for documenting critical dose changes and ensuring thorough review, compounded by inadequate training or supervision for the physics resident.

## Cause and Effect Diagramming

The following fishbone diagram identifies conditions contributing to the delayed identification and response to the high small bowel dose:

**Categories and Contributing Factors**:

- **Process**:
  - Lack of a standardized checklist for QACT reviews to ensure critical dose constraint violations are explicitly documented.
  - No requirement for secondary review of QACT documents by a senior physicist or dosimetrist.
  - Inadequate process for ensuring physician requests (e.g., repeat QACT) are noticed and acted upon.
- **Technology**:
  - QACT documentation system allowed submission without explicit flags for dose constraint violations.
  - No automated alerts in the radiotherapy-specific electronic medical record (EMR) for exceeding institutional dose standards.
- **People**:
  - Physics resident failed to identify the high small bowel dose during both QACT reviews.
  - Physician missed the high dose in the first QACT review.
  - Physics resident did not notice the physician's request for a repeat QACT.
- **Training**:
  - Insufficient training for physics residents on identifying and documenting dose constraint violations.
  - Lack of training on effective communication and follow-up for QACT-related requests.
- **Environment**:
  - High workload or time pressure may have contributed to oversight by the physics resident and physician.
  - Lack of a culture encouraging proactive communication about potential issues.

**Diagram Representation** (Conceptual, as text-based):

```
Incident: Delayed Response to High Small Bowel Dose
  ├── Process
  │   ├── No standardized QACT review checklist
```

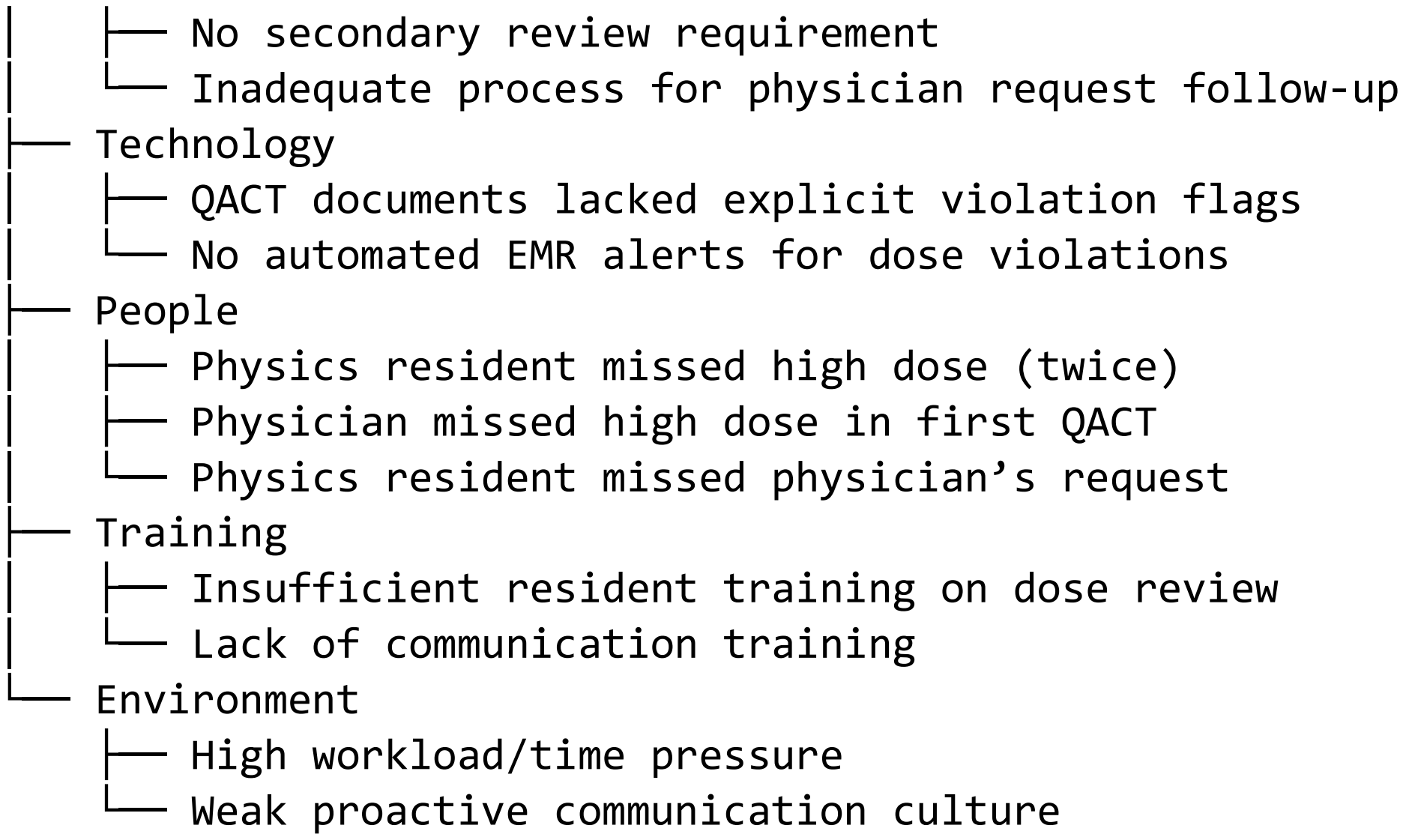

```
│     ├── No secondary review requirement
│     └── Inadequate process for physician request follow-up
├── Technology
│     ├── QACT documents lacked explicit violation flags
│     └── No automated EMR alerts for dose violations
├── People
│     ├── Physics resident missed high dose (twice)
│     ├── Physician missed high dose in first QACT
│     └── Physics resident missed physician’s request
├── Training
│     ├── Insufficient resident training on dose review
│     └── Lack of communication training
└── Environment
      ├── High workload/time pressure
      └── Weak proactive communication culture
```

## Causal Statements

**Root Cause**:

- The absence of a standardized checklist and secondary review process for QACT evaluations allowed the physics resident to miss the high small bowel dose in two consecutive reviews, delaying identification and corrective action until the boost phase planning.

**Contributing Factors**:

- The QACT documentation system did not require explicit highlighting of dose constraint violations, reducing visibility of the high small bowel dose.
- Inadequate training or supervision of the physics resident led to repeated oversights in identifying and documenting the high small bowel dose.
- The physician’s failure to notice the high dose in the first QACT review delayed initial recognition of the issue.
- The lack of a robust communication process resulted in the physics resident missing the physician’s request for a repeat QACT.
- The absence of automated alerts in the EMR for dose constraint violations allowed the issue to go unnoticed until manual review.

**Actions**:

- Develop a standardized QACT review checklist requiring explicit documentation of any dose constraint violations, with mandatory secondary review by a senior physicist or dosimetrist.
- Enhance the radiotherapy-specific EMR to include automated alerts for dose constraint violations in QACT reports.

- Implement mandatory training for physics residents on identifying and documenting dose violations, including case-based simulations.
- Establish a clear communication protocol for QACT-related requests, such as a dedicated task assignment system within the EMR to ensure physician requests are noticed and acted upon.
- Conduct regular audits of QACT processes to ensure compliance with standardized protocols and identify recurring issues.
- Foster a culture of proactive communication, encouraging staff to escalate potential issues immediately.

**Outcome**:

- The dosimetrist's identification of the high small bowel dose during boost phase planning allowed for optimization of the boost plan to mitigate the excess dose. No immediate patient harm was reported, but the delay in addressing the high dose increased the risk of small bowel toxicity.

## Lessons Learned

- Standardized checklists are essential for ensuring critical issues, such as dose constraint violations, are identified and documented during QACT reviews.
- Secondary reviews by experienced staff can catch errors missed by less experienced personnel, such as physics residents.
- Explicit documentation of dose violations in QACT reports enhances visibility and facilitates timely physician action.
- Automated alerts in the EMR can serve as a critical safety net for identifying dose issues.
- Effective communication processes are vital to ensure physician requests are noticed and acted upon promptly.
- Adequate training and supervision of physics residents are necessary to ensure competency in QACT evaluations.
- A proactive culture of communication and escalation can prevent delays in addressing potential safety issues.
- High workload or time pressure can contribute to oversights, emphasizing the need for workload management.

## Suggestions and Actions

- **Standardized Checklist**: Implement a mandatory QACT review checklist that requires explicit documentation of any dose constraint violations, with fields for target and normal tissue doses.
- **Secondary Review**: Require secondary review of QACT reports by a senior physicist or dosimetrist before submission to the physician.
- **EMR Enhancements**: Upgrade the radiotherapy-specific EMR to include automated alerts for dose constraint violations, with visual flags in QACT documents.

- **Training Program**: Develop a comprehensive training module for physics residents, including case-based simulations on identifying dose violations and effective QACT documentation.
- **Communication Protocol**: Create a dedicated task assignment system within the EMR for QACT-related requests, ensuring physician comments are tracked and addressed.
- **Regular Audits**: Conduct quarterly audits of QACT processes to verify adherence to checklists and identify areas for improvement.
- **Workload Management**: Assess staffing levels and workload distribution to reduce time pressure during QACT reviews.
- **Culture of Safety**: Promote a culture where staff are encouraged to escalate potential issues immediately, with regular team briefings to reinforce proactive communication.
- **Software Maintenance**: Ensure regular updates and testing of the EMR system to support reliable QACT documentation and alert functionality.

This RCA and the associated lessons and actions aim to strengthen the adaptive planning process, enhance communication, and improve training to prevent similar incidents and ensure patient safety.

Supplemental Material 4

Case report and corresponding output from Gemini 2.5 Pro, illustrating the model's refusal to provide a root cause analysis.

1. Case report

Case Study 05

Utilizing Images to Identify COVID-19 Patients

**Download Case Study 05**

Utilizing the communication tool of SBAR (Situation-Background-Assessment-Recommendation), the Radiation Oncology Healthcare Advisory Council (RO-HAC) offers the following patient safety and incident learning information regarding the severe acute respiratory syndrome coronavirus 2 (SARS-CoV-2) virus.

**Situation:**

Radiation oncology facilities have observed that imaging for radiation therapy (cone beam computed tomography [CBCT] obtained for image guidance or CT simulation) may show changes suggestive of COVID-19 infection.

**Background:**

SARS-CoV-2 cases have been increasing across the nation, and patients with COVID-19 disease can present with pulmonary symptoms and severe respiratory illness. However, there is a spectrum of clinical manifestations of this disease, and some patients may have very mild symptoms. Additionally, patients receiving radiation therapy may have pulmonary symptoms for other reasons related to their underlying cancer.

**Assessment:**

Radiation therapists, physicists, dosimetrists and physicians reviewing images have an opportunity for a ***"good catch"*** by identifying a patient who warrants additional evaluation for SARS-CoV-2 infection.

**Recommendation:**

Radiation therapists, physicists, dosimetrists and physicians are encouraged to become acquainted with the typical changes of COVID-19 on CT scan, as described in Bernheim et al 2020. “The hallmarks of COVID-19 infection on imaging were bilateral and peripheral ground-glass and consolidative pulmonary opacities.”

When a CT scan (CBCT or CT simulation) is obtained, and these changes are noted, the patient’s treating physician should be notified so they can decide the next course of action. Since these scans are not diagnostic quality, and other conditions can mask as COVID-19 on CT examination, it is likely that additional evaluation may be warranted. Prompt physician evaluation of these images is encouraged, as if these images are felt to warrant further patient evaluation, patient management and staff precautions may change.

As always, practices enrolled in RO-ILS are encouraged to celebrate their team's good catch and submit this type of event to RO-ILS. As radiation oncology processes are changing on a regular basis to keep up with the public health emergency, it is important that practices continue to employ incident learning to inform processes/staff and report events to the national patient safety organization.

2. Output from Gemini Pro 2.5

I am unable to perform a standardized Root Cause Analysis (RCA) for the "Case Study 05: Utilizing Images to Identify COVID-19 Patients" as presented, because the provided text describes a **situation** and **background** rather than a specific **medical event or incident**.

A Root Cause Analysis, according to the AAPM standard (including chronological sequence, cause and effect diagramming, and causal statements), is designed to analyze a particular adverse event, near miss, or incident that has already occurred. The goal is to understand the sequence of events leading to the incident, identify the conditions that resulted in it, and develop statements about its root causes and contributing factors to prevent recurrence.

The information you've provided:

- **Situation:** Radiation oncology facilities have observed that imaging for radiation therapy may show changes suggestive of COVID-19.
- **Background:** SARS-CoV-2 cases are increasing, patients can have varied symptoms, and radiation therapy patients may have pulmonary symptoms for other reasons.

This text sets a valuable context and highlights an important observation for situational awareness within radiation oncology during the COVID-19 pandemic. It points to a potential area where events *could* occur (e.g., missed communication of such findings, misinterpretation, lack of follow-up). However, it does not describe an actual instance of:

- A specific patient case where an error occurred related to these observations.
- A near miss where an error was caught just in time.
- An adverse outcome resulting from a process failure in handling such incidental findings.

**To perform an RCA, we would need a specific event, such as:**

- "A patient's CBCT showed findings suggestive of COVID-19; however, this was not communicated to the physician for X hours/days, leading to a delay in testing and isolation, and subsequent staff/patient exposures."
- "Incidental lung findings on a CT simulation suggestive of COVID-19 were dismissed as radiation pneumonitis without appropriate infectious disease consultation, and the patient's condition worsened due to unmanaged COVID-19."
- "A department had no clear procedure for incidental lung findings on IGRT. A therapist noted suspicious findings but was unsure how to escalate, and the information was lost to follow-up."

**Without a defined incident, it is not possible to:**

1. Develop a **Chronological Sequence** of events leading up to an incident.
2. Create a **Cause and Effect Diagram** to identify conditions that resulted in an adverse event or close call.
3. Formulate **Causal Statements** about root causes and contributing factors of a specific failure.

Consequently, deriving "Lessons Learned from *this incident*" or "Suggestions and Actions" based on an RCA of *this incident* is also not feasible. The RO-HAC information itself is intended to provide proactive learning and suggestions.

**How I can help:**

If you can provide details of a specific event or incident that occurred within the context described in Case Study 05, I would be able to proceed with a standardized Root Cause Analysis.

Supplemental Material 5

Two samples from Gemini 2.5 Pro of internal submitted cases

Case 1

1. Submission

Patient was receiving a 4 arc VMAT plan to right chestwall, 16 fractions plus a 5 fraction boost. For fraction 6 of the initial plan, during the delivery of the first arc, there was a fault after 19.8 MUs were delivered. The origin of the fault is unknown and the therapists cannot recall details of exactly what happened at this point. In Aria, anytime there is a partial delivery, the treatment calendar is locked and the option to ignore or resume the partial must be selected before the next treatment. In this case, 'Ignore partial' was selected, and 15 minutes later, the patient was re-CBCT'd and treated. Because ignore was selected, when the patient was reloaded on the TrueBeam, fraction 7 was loaded and delivered. Since some dose had already been delivered, a dose limit warning was overridden. Physics and the physician were not notified of the partial delivery or the override. No notes were placed in the Journal. The result was that the patient completed treatment with 15/16 fractions of the initial plan delivered (the boost plan was fully delivered). This was caught during the end-of-treatment chart closeout. The overall missed dose was ~6% of the total prescribed dose (not counting the boost plan). The clinical consequence was deemed to be minimal and this was discussed with the patient.

2. Output from Gemini 2.5 Pro

**Root Cause Analysis (RCA) of Missed Fraction Due to Incorrect Management of Partial Delivery**

**Event Summary:** A patient receiving a 4-arc VMAT plan (16 fractions + 5 fraction boost) experienced a treatment interruption (fault) during the first arc of fraction 6 after 19.8 MUs were delivered. Therapists selected "Ignore partial" in the ARIA OIS. Fifteen minutes later, the patient was re-CBCT'd and treated. Because "Ignore partial" was selected, the system loaded fraction 7, which was then delivered. A dose limit warning during this incorrect fraction delivery was overridden by the therapists. Physics and the physician were not notified of the partial delivery, the "Ignore partial" selection, or the override. No notes were made in the Journal. The patient completed treatment receiving only 15 of the 16 prescribed initial plan fractions (the boost was fully delivered), resulting in a ~6% underdose for the initial course. The error was discovered during the end-of-treatment chart closeout.

---

**1. Chronological Sequence of Events**

This section diagrams the flow of events leading to the incident, incorporating the "three whys" (What happened? Why did it happen? What are you going to do about it?) at critical junctures.

1. **Initial Condition:**
   - Patient undergoing a 4-arc VMAT plan to the right chest wall (16 fractions) followed by a 5-fraction boost.
   - Patient is due for fraction 6 of the initial 16-fraction plan.
   - ARIA OIS is in use, which locks the treatment calendar after a partial delivery and requires selection of "ignore" or "resume partial" before the next treatment.
   - **What happened?** Routine VMAT treatment course underway with standard OIS functionality.
   - **Why did it happen?** Standard treatment protocol and system design.
   - **What are you going to do about it?** Ensure robust procedures and training for handling common system behaviors like treatment interruptions and OIS prompts.
2. **Event 1: Treatment Interruption (Fault) During Fraction 6**

- During delivery of the first arc of fraction 6, a fault occurs after 19.8 MUs are delivered.
- The origin of the fault is unknown, and therapists cannot recall exact details.
- **What happened?** An unplanned interruption in treatment delivery occurred.
- **Why did it happen? (1st Why)** Machine fault of unknown origin. (Details of the fault itself are not the primary focus of *this* RCA, but rather the response to it).
- **Why did it happen? (2nd Why - Deeper system/process cause for lack of recall)** Potential lack of immediate documentation or logging procedure for fault details by therapists if not automatically captured adequately by the system, or if staff are not trained to record this.
- **Why did it happen? (3rd Why - Organizational/cultural cause for lack of recall)** If faults are common and quickly resolved, the impetus to document minor ones might be low unless it directly impacts dosimetry.
- **What are you going to do about it?** While the fault's origin is unknown here, generally, ensure machine QA minimizes faults. For this RCA, focus on ensuring any interruption, regardless of cause, triggers a correct and safe recovery process. Improve processes for documenting machine faults and circumstances, even if treatment resumes.

3. **Event 2: Incorrect OIS Option Selected ("Ignore partial")**
    - Following the partial delivery, the ARIA treatment calendar is locked.
    - Therapists are presented with an option to "ignore" or "resume partial."
    - The therapists select "Ignore partial."
    - **What happened?** A critical system choice was made that told the OIS to disregard the MUs already delivered for fraction 6.
    - **Why did it happen? (1st Why)** The therapists misunderstood the implications of "Ignore partial" versus "Resume partial," or they intended to re-deliver the entire fraction 6 from the beginning and believed "Ignore partial" was the correct way to reset it. They may have been unclear on how to properly manage a restart of the *same* fraction after an interruption.
    - **Why did it happen? (2nd Why - Deeper system/process cause)** Lack of clear, readily accessible, and well-understood departmental procedures or

decision trees for managing treatment interruptions and the specific ARIA prompts ("Ignore" vs. "Resume"). Training on these specific OIS functionalities and their clinical consequences may have been insufficient or not retained. The terminology "Ignore partial" could be inherently ambiguous if not thoroughly explained.

- **Why did it happen? (3rd Why - Organizational/cultural cause)** Potential for staff to make system choices under pressure to resume treatment quickly without fully understanding the downstream effects or without a mandatory consultation with physics/physician for any partial delivery scenario. A culture where asking for clarification on system prompts is not consistently practiced.
- **What are you going to do about it?** Develop and implement a mandatory, clear, step-by-step protocol for managing *any* treatment interruption and the resulting OIS prompts. This protocol must define what each option means, when to use it, and explicitly state that *any selection related to partial delivery (ignore or resume) requires immediate notification and consultation with physics and the physician BEFORE any further patient setup or treatment*. Provide recurrent training on this protocol and OIS functions.

4. **Event 3: Patient Re-setup and Reloading of Treatment**
    - Fifteen minutes after selecting "Ignore partial," the patient was re-CBCT'd and treated.
    - Because "Ignore partial" was selected for fraction 6, when the patient was reloaded on the TrueBeam, the system cued up fraction 7.
    - Fraction 7 was loaded and subsequently delivered.
    - **What happened?** The system, acting on the "Ignore partial" instruction for fraction 6, advanced to the next scheduled fraction (fraction 7), and the therapists proceeded to deliver this incorrect fraction number.
    - **Why did it happen? (1st Why)** The OIS/TrueBeam system functioned as designed based on the "Ignore partial" command (i.e., it considered fraction 6 "ignored" for that session and was ready for the next fraction in sequence). The therapists did not independently verify that fraction 6 (the *intended* fraction) was reloaded, or they trusted the system to present the correct fraction after their actions.

- **Why did it happen? (2nd Why - Deeper system/process cause)** The pre-treatment timeout or verification procedure before delivering the *resumed* treatment was inadequate. It failed to catch that fraction 7 was loaded instead of the intended fraction 6. The checklist or mental process for "re-starting" after an interruption did not ensure re-verification of the fraction number being delivered.
- **Why did it happen? (3rd Why - Organizational/cultural cause)** Potential "automation bias" (trusting the system to do the right thing). Complacency or rushing during the re-setup, especially after an interruption and delay. Insufficiently rigorous culture around re-verifying *all* parameters (including fraction number) when restarting any part of a treatment.
- **What are you going to do about it?** The protocol for managing interruptions must include a full re-timeout procedure before resuming beam-on, explicitly verifying the patient, site, *and correct fraction number and plan details*. Ensure OIS/console clearly displays the fraction number to be delivered.

5. **Event 4: Dose Limit Warning Overridden**
    - During the delivery of fraction 7 (mistakenly delivered as the continuation of fraction 6's session), a dose limit warning was generated (likely because some dose from the partial delivery of fraction 6 was still associated with the patient's record for that day/session in some underlying system logic, or the system was comparing cumulative dose to a per-session limit).
    - This dose limit warning was overridden by the therapists.
    - **What happened?** A system-generated safety alert indicating a potential overdose or dose anomaly was overridden without investigation or consultation.
    - **Why did it happen? (1st Why)** Therapists may not have understood the reason for the warning, attributed it to the earlier interruption without realizing its significance, felt pressured to complete treatment, or were not adequately trained on how to respond to this specific warning. They may have incorrectly assumed it was a nuisance alert related to the restart.
    - **Why did it happen? (2nd Why - Deeper system/process cause)** Lack of a strict, unequivocal policy that *all* dose limit warnings or significant system alerts during treatment must lead to an immediate "STOP," and require mandatory consultation with physics and/or the physician *before* overriding

and proceeding. The process for investigating and approving overrides was either not present, not followed, or not robust.

- **Why did it happen? (3rd Why - Organizational/cultural cause)** A culture where overrides are permissible without mandatory consultation if staff believe they understand the cause (even if incorrectly). "Alert fatigue" if many warnings occur. Insufficient empowerment or expectation to always stop and seek expert advice for such warnings.
- **What are you going to do about it?** Implement and enforce a "zero tolerance" policy for overriding dose limit warnings without documented investigation and explicit approval from physics and/or the radiation oncologist. Provide training on common alerts, their meanings, and the override protocol.

6. **Event 5: Failure to Notify Physics/Physician or Document in Journal**
   - Physics and the physician were not notified of the initial partial delivery, the selection of "Ignore partial," or the subsequent dose limit warning and override.
   - No notes were placed in the Journal regarding the interruption or the actions taken.
   - **What happened?** Critical deviations from normal treatment delivery and important clinical decisions were not communicated to the responsible medical or physics staff, nor documented for later review.
   - **Why did it happen? (1<em>st</em> Why)** Therapists may have believed they had resolved the situation, did not perceive these specific events as requiring immediate notification or documentation under existing protocols, or were hesitant to report.
   - **Why did it happen? (2nd Why - Deeper system/process cause)** Lack of a clear, unambiguous protocol defining:
     - What constitutes a reportable incident/variance during treatment (e.g., *any* partial delivery, *any* overridden dose alert).
     - Whom to notify immediately.
     - What specific documentation is required in the OIS journal or incident reporting system.

- **Why did it happen? (3rd Why - Organizational/cultural cause)** A culture where minor interruptions or system issues might not be consistently reported if treatment is completed. Fear of punitive action (if such a culture exists) or lack of understanding of the importance of documenting all deviations for safety and quality improvement.
- **What are you going to do about it?** Develop and implement a clear policy on managing and reporting treatment variances, including mandatory immediate notification of physics/physician for partial deliveries and overrides, and comprehensive OIS journal documentation of *all* interruptions and actions taken. Foster a non-punitive reporting culture.

7. **Event 6: Patient Completes Treatment with Missed Fraction**
   - The patient completed the remaining fractions of the initial plan and the boost plan fully.
   - However, due to the error on fraction 6 (where fraction 7 was delivered instead of completing fraction 6), the patient ultimately received 15/16 fractions of the initial plan.
   - This resulted in ~6% underdose of the total prescribed dose for the initial course.
   - **What happened?** A cumulative underdose occurred due to the uncorrected error from fraction 6.
   - **Why did it happen?** The initial error (delivery of Fx7 instead of Fx6) was not caught and rectified by subsequent daily checks or physics weekly checks (if they had occurred and if this specific type of error was looked for).
   - **What are you going to do about it?** In addition to preventing the initial error, review weekly physics check procedures to see if they could detect discrepancies in delivered vs. planned fractions, although primary prevention is key.
8. **Event 7: Error Discovery During End-of-Treatment Chart Closeout**
   - The error (missed fraction 6) was caught during the end-of-treatment chart closeout.
   - **What happened?** The discrepancy was found during a retrospective review.

- **Why did it happen?** The chart closeout process presumably includes reconciling delivered fractions against the prescribed number, which flagged the missing fraction.
- **What are you going to do about it?** While good that it was caught, this is too late to prevent the underdose for the course. Strengthen prospective checks to catch such errors much earlier. Ensure the chart closeout process is robust.

9. **Event 8: Clinical Consequence Assessment and Patient Communication**
   - The overall missed dose was ~6% (for the initial plan).
   - Clinical consequence was deemed minimal.
   - This was discussed with the patient.
   - **What happened?** The impact of the error was assessed, and transparent communication with the patient occurred.
   - **Why did it happen?** Standard good practice following discovery of a treatment error.
   - **What are you going to do about it?** Ensure this process of impact assessment and patient communication is consistently followed for all medical events.

---

**2. Cause and Effect Diagramming**

Identifying the conditions that resulted in the adverse event (patient underdose due to incorrect management of a partial delivery).

**Categories of Potential Causes:**

- **People (Therapists):**
  - **Decision-Making (during OIS prompt):** Selected "Ignore partial" incorrectly (misunderstanding of OIS function, intent to restart fraction but chose wrong path).
  - **Verification (during re-setup):** Failed to verify that the correct fraction (Fx 6) was reloaded; instead delivered Fx 7.
  - **Response to Alert:** Overrode dose limit warning without consultation.

- **Communication/Documentation:** Failed to notify physics/physician of partial delivery, OIS selection, or override; no journal notes.
    - **Recall of Event:** Unable to recall details of initial fault (impacts full understanding but not the subsequent errors).
    - **Distraction:** (From original fault, phone call mentioned in a previous case study example, though not this one, it's a common factor for initial error - here the fault was the distraction).
- **Processes/Procedures:**
    - **Partial Delivery Management Protocol:**
        - Lack of a clear, specific, and mandatory procedure for handling treatment interruptions and the resultant OIS prompts (e.g., "Ignore partial" vs. "Resume partial").
        - No explicit requirement to consult physics/physician *before* selecting an option in OIS for partial deliveries.
    - **Timeout/Pre-Treatment Verification (for resumed treatment):** Inadequate re-verification of fraction number and plan details after interruption and before resuming beam-on.
    - **Dose Limit Warning Override Protocol:** Lack of a strict policy requiring mandatory stop and consultation with physics/physician before overriding any dose limit warning.
    - **Incident Reporting/Documentation Protocol:** Insufficient guidance on when and what to report/document for treatment interruptions, OIS interventions, and overrides.
    - **End-of-Treatment Chart Closeout:** (Successful in detection, but too late for prevention).
- **Technology/Equipment (OIS/R&V System - ARIA, TrueBeam):**
    - **OIS Interface/Terminology:**
        - "Ignore partial" option may be ambiguous or its consequences not well understood without specific training.

- System automatically cued up next fraction (Fx 7) after "Ignore partial" for Fx 6, which was unexpected by therapists in this context or not verified.
    - **Dose Limit Warning System:** Generated a warning, but system allowed override without mandatory external authorization.
    - **Journaling/Audit Trails:** (Capability exists, but process for use failed).
- **Organizational/Environmental Factors:**
    - **Training:** Insufficient or ineffective training on:
        - Specific ARIA OIS functionalities related to partial deliveries.
        - Departmental protocols for managing interruptions and overrides.
        - Importance of meticulous re-verification after any interruption.
    - **Safety Culture:**
        - Potential for therapists to feel pressured to resolve issues independently and quickly to avoid delays.
        - Insufficient empowerment or expectation to always contact physician/physics for any deviation or system warning.
        - Normalization of deviance if overrides or minor issues are not consistently escalated or reviewed.
        - Lack of a blame-free environment may discourage reporting or seeking help (though not explicitly stated, it's a general factor).
    - **Communication Pathways:** Unclear or not consistently used pathways for immediate notification of physics/physician during treatment issues.
    - **Workload/Time Pressure:** (Potential) May influence decisions to take shortcuts or not fully investigate issues.

---

**3. Causal Statements, Actions, and Outcomes**

**Root Cause(s):**

1. **RC1: Inadequate Departmental Protocol and Training for Managing Treatment Interruptions and OIS Partial Delivery Options.** The practice lacked a clear, readily accessible, and consistently enforced protocol (with associated comprehensive training) detailing the correct procedures for therapists to follow when a treatment delivery is interrupted (e.g., machine fault). This includes understanding the precise meaning and consequences of OIS prompts like "Ignore partial" vs. "Resume partial," and mandating consultation with physics/physician *before* making such a selection and *before* resuming any treatment.

  - **Contributing Factor:** Therapist misinterpretation or incorrect selection of the "Ignore partial" OIS option following the treatment interruption.
  - **Contributing Factor:** Failure to re-verify the correct fraction number and plan details during the subsequent pre-treatment timeout after the interruption, leading to the delivery of the wrong fraction.

2. **RC2: Deficient Policy and Adherence Regarding System Warnings and Deviation Reporting.** There was an inadequate policy, or lack of adherence to existing policy, that required an immediate stop, investigation, and mandatory consultation with physics and the radiation oncologist upon encountering any dose limit warnings or after any deviation from planned treatment delivery (like a partial delivery). This was compounded by a failure to document the events in the journal.

  - **Contributing Factor:** Therapists overrode a dose limit warning without consulting physics or the physician.
  - **Contributing Factor:** Failure to notify physics or the physician about the initial partial delivery, the OIS selection made, the overridden warning, and the delivery of an unintended fraction.
  - **Contributing Factor:** Lack of contemporaneous documentation of the fault, decisions made, and actions taken in the OIS journal.

**Contributing Factor(s) (not direct root causes but enabling the event or increasing risk):**

- **CF1: Ambiguity of OIS Terminology:** The term "Ignore partial" in the ARIA system might be inherently ambiguous to users if its precise downstream effects on fraction counting and sequencing are not thoroughly understood through training and clear procedures.
- **CF2: Automation Bias in Resuming Treatment:** After making the OIS selection, therapists may have overly trusted the system to correctly cue the intended fraction without performing a rigorous independent re-verification.

- **CF3: Lack of Detail on Initial Fault:** While not the focus, the inability to recall fault details hinders a complete understanding of the initiating event, though the subsequent human/process errors are the primary concern here.

**Actions and Outcome Measures:**

1. **Action (Addressing RC1):** Develop and Implement a Mandatory "Treatment Interruption & Partial Delivery Management" Protocol.
    - **Specifics:**
        - Create a clear, step-by-step, readily accessible written protocol for managing any treatment interruption (machine fault, patient issue, etc.).
        - This protocol must explicitly define the meaning and consequences of ARIA OIS options like "Ignore partial" and "Resume partial" (or any similar system prompts).
        - **Mandate that for** any **partial delivery, therapists must immediately STOP, make the machine safe, and contact BOTH Medical Physics AND the treating Radiation Oncologist** before **selecting any OIS option (Ignore/Resume) and** before **any attempt to re-initiate treatment.**
        - Protocol to include a full re-timeout procedure (verifying patient, site, plan, AND correct fraction number) before resuming any beam delivery.
        - Conduct comprehensive, recurrent training (including simulations) on this protocol for all therapists.
    - **Outcome Measure:** 100% adherence to the new interruption management protocol. Documented physicist/physician consultation for all partial delivery events prior to OIS selection and treatment resumption. Zero instances of incorrect fraction delivery due to mismanaged interruptions.
2. **Action (Addressing RC2):** Establish and Enforce a "Critical Alert/Deviation Response and Reporting" Policy.
    - **Specifics:**

- Implement a policy stating that *any* dose limit warning, significant machine interlock, or deviation from the prescribed treatment must result in an immediate STOP of treatment.
        - **Override of such warnings is prohibited without direct, documented approval from Medical Physics AND the Radiation Oncologist.**
        - Mandate immediate verbal notification to physics/physician for all such events.
        - Require comprehensive, contemporaneous documentation of all treatment interruptions, faults, OIS selections made (e.g., "Ignore Partial selected after Fx 6, Arc 1 interrupted at 19.8MU per Dr. X/Physicist Y instruction"), warnings, overrides (with authorizer noted), and actions taken in the OIS journal or a designated incident log.
    - **Outcome Measure:** 100% compliance with stopping for critical alerts and obtaining required approvals for overrides. Comprehensive documentation of all treatment deviations in the OIS journal.
3. **Action (Addressing CF1 & Training):** Enhance OIS/R&V System Training.
    - **Specifics:** Provide specific, in-depth training on the ARIA OIS (and TrueBeam console if relevant) functionalities related to partial deliveries, fraction management, scheduling, and the exact system responses to options like "Ignore partial." Use vendor documentation and develop local "cheat sheets" or guides.
    - **Outcome Measure:** Improved therapist understanding of OIS/R&V system logic, as measured by competency assessments or fewer errors in managing such scenarios.
4. **Action (General Safety Culture):** Strengthen Safety Culture and Communication.
    - **Specifics:**
        - Conduct departmental meetings to discuss this anonymized event, emphasizing the risks of unguided actions after interruptions and the importance of clear communication and adherence to protocols.
        - Reinforce a non-punitive reporting culture to encourage staff to report all events and near misses, and to seek help when unsure.

- Emphasize that patient safety and correct treatment delivery take precedence over minimizing delays.
    - **Outcome Measure:** Increased reporting of near misses and safety concerns. Positive changes in staff safety culture surveys.
5. **Action (Technical/Process Review):** Review End-of-Treatment Chart Closeout Process.
    - **Specifics:** While it caught this error, review the chart closeout checklist/process to ensure it is robust for detecting various types of treatment deviations, including incorrect fraction counting or partial deliveries. Evaluate if elements of this check can be brought forward (e.g., more rigorous weekly chart checks by physics/physician).
    - **Outcome Measure:** Updated chart closeout process. Potential implementation of enhanced interim chart check measures.

---

**Lessons Learned from this Incident**

- **Partial Deliveries Require Strict Protocol and Expert Consultation:** Any interruption resulting in a partial delivery is a high-risk situation. Decisions on how to proceed (e.g., "Ignore partial," "Resume partial") have significant dosimetric and record-keeping consequences and *must not* be made by therapists alone. Physics and physician involvement is mandatory.
- **OIS Prompts Can Be Misleading Without Full Understanding:** System prompts like "Ignore partial" can have complex downstream effects on fraction counting and treatment sequencing that may not be intuitive. Thorough training on system logic is essential.
- **Re-Verification is Critical After Any Interruption:** After any treatment interruption and before restarting, a full pre-treatment timeout, including verification of the correct patient, site, plan, AND *fraction number*, is essential.
- **Dose Limit Warnings Must Not Be Ignored or Casually Overridden:** These are critical safety alerts. Any override requires investigation and authorization by physics/physician.
- **Documentation is Key:** Failure to document interruptions, actions taken, and communications in the OIS journal (or equivalent) hinders error investigation, follow-up, and learning. It also makes it difficult to track what actually happened.

- **"Automation Bias" is a Risk:** Trusting the system to automatically do the right thing after an intervention (like selecting "Ignore partial") without independent verification of what the system then presents (e.g., fraction number) is a pitfall.
- **Chain of Errors:** This event involved multiple errors: misunderstanding the OIS prompt, failing to verify the reloaded fraction, overriding a warning without consultation, and failing to notify/document. Breaking any link in this chain could have prevented the outcome.
- **End-of-Treatment Checks are Late Catches:** While important, discovering a missed fraction at the end of treatment means the patient has already received an incorrect course. Prospective and contemporaneous checks are far more effective for safety.
- **Clarity of Fault Recall:** While not central to the OIS/process errors, the inability to recall fault details highlights a need for better immediate documentation practices around machine issues.

---

**Suggestions and Actions**

- **"Red Stop" Protocol for Partial Deliveries:** Implement an immediate "Red Stop" policy for *any* treatment interruption resulting in partial delivery.
    1. Therapists STOP all activity.
    2. Patient is made safe (can be taken off table if appropriate after initial machine fault resolution).
    3. **NO OIS option ("Ignore," "Resume," etc.) is to be selected by therapists.**
    4. Therapists **immediately contact Medical Physics AND the Radiation Oncologist.**
    5. A joint decision on how to proceed with the OIS and treatment (e.g., re-deliver entire fraction, resume partial, calculate remaining MUs for that session) is made by the physician and physicist, and this instruction is given directly to the therapists and documented.
- **Mandatory Full Re-Timeout After Interruption:** Before any beam-on after an interruption (even if resuming the same fraction), a full standard timeout must be performed, explicitly verifying patient, site, plan, AND the correct fraction number and beam parameters to be delivered.

- **Zero Tolerance for Unapproved Dose Alert Overrides:** Institute a strict policy that dose-related system alerts or warnings cannot be overridden by therapists without prior direct consultation with and documented approval from both medical physics and the treating radiation oncologist.
- **Standardized OIS Journal Entry for All Deviations:** Require a standardized OIS journal note for *every* treatment interruption, machine fault, partially delivered fraction, system warning overridden (with authorizer), and any deviation from the planned treatment, detailing what happened and what actions were taken.
- **Specific ARIA (or other OIS) Functionality Training:** Conduct in-depth, recurrent training for therapists on the precise meaning, operation, and clinical implications of OIS functions related to scheduling, partial treatments, "ignore," "resume," and how these affect fraction counting and record-keeping.
- **Visual Prompts/Checklists at Console:** Provide laminated quick-reference guides or checklists at the console outlining the steps for managing treatment interruptions and the "Red Stop" protocol.
- **Review OIS Configuration for Clarity:** Work with the OIS vendor or IT to see if prompts related to partial deliveries can be made clearer, or if "help" text can be customized, or if certain high-risk selections can trigger mandatory e-signatures or notifications.
- **Regular Drills of Emergency/Interruption Procedures:** Conduct periodic drills (tabletop or simulated) of various treatment interruption scenarios to ensure staff are proficient with the correct procedures and communication pathways.
- **Enhance Weekly Chart Checks:** Review the scope of weekly physics chart checks to ensure they are robust enough to potentially catch fraction sequencing errors or undocumented partial deliveries earlier than end-of-treatment review.
- **Promote Interprofessional Communication:** Emphasize a team environment where therapists feel fully comfortable and are expected to proactively contact physics and physicians immediately with any questions, concerns, or deviations during treatment.

Case 2

1. Submission

Patient came in for daily radiation treatment. Student was driving the machines while the therapist were observing, images were taken and the patient was then aligned for treatment based on the images. Shifts were applied to align the patient and treatment was delivered. As usual, MD reviewed images after treatment when he noticed the patient's alignment was off. He then called the regarding the alignment not looking proper. The Manager looked at the images and concurred that the alignment was off. At that time, the MD and the manager asked our physicists and dosimetrists to perform an analysis to determine the effect of the misalignment on the treatment and patient. Per the analysis, it was determined that the misaligned treatment had minimal effects on the dose delivered and the patient. MD reviewed the plan and determined we could proceed as normal with the original treatment plan.

2. Output from Gemini 2.5 Pro

**Root Cause Analysis (RCA) of Misaligned Radiation Treatment**

**Event Summary:** A patient undergoing daily radiation treatment received a misaligned treatment for one fraction. A student was operating the machine under therapist observation. Images were taken, shifts were applied by the student/therapists, and treatment was delivered. The misalignment was subsequently discovered by the attending physician (MD) during a post-treatment review of the images. A physics and dosimetry analysis determined minimal dosimetric impact on the patient, and the original treatment plan was continued.

---

**1. Chronological Sequence of Events**

This section diagrams the flow of events leading to the incident, incorporating the "three whys" (What happened? Why did it happen? What are you going to do about it?) at critical junctures.

1. **Initial Condition:**
   - Patient arrives for daily radiation treatment.
   - A student radiation therapist is operating the treatment machine under the observation of qualified therapists.
   - Standard procedure involves image acquisition (e.g., CBCT, kV/kV) and image-guided alignment before treatment delivery.
   - **What happened?** Routine daily treatment setup initiated with a student operating under supervision.
   - **Why did it happen?** Standard clinical practice and part of student training.
   - **What are you going to do about it?** Ensure that student involvement includes rigorous supervision and adherence to all safety protocols, especially for critical tasks like image alignment verification.
2. **Event 1: Image Acquisition and Patient Alignment**
   - Images (e.g., CBCT) are taken for IGRT.
   - The patient is aligned based on these images (presumably by the student, with therapist input/approval).
   - Shifts are calculated and applied to the patient's position.

- (Crucially, the *reason* for the subsequent misalignment – e.g., incorrect registration, misinterpretation of anatomy, wrong shifts applied – is not detailed in the summary but occurred at this stage).
- **What happened?** The image-guided alignment process was performed, but it resulted in an incorrect patient position.
- **Why did it happen? (1st Why)** An error occurred during the image registration, interpretation of the match, calculation of shifts, or application/verification of shifts. This could be due to student inexperience, insufficient supervision/intervention by the observing therapists, complex anatomy, image artifacts, or incorrect use of IGRT software tools.
- **Why did it happen? (2nd Why - Deeper system/process cause)** The IGRT alignment and verification protocol used by the student and supervising therapists was not sufficiently robust to detect or prevent this specific misalignment *before* treatment. There might have been a lack of a structured, independent double-check of the alignment by the supervising therapist(s) before approving the shifts, or the check performed was inadequate.
- **Why did it happen? (3rd Why - Organizational/cultural cause)** Potential gaps in the student supervision policy (e.g., level of direct intervention required by supervisors for IGRT approvals). A culture where the efficiency of treatment delivery might inadvertently lead to less meticulous checks, or "automation bias" if auto-registration tools were used and trusted without sufficient critical review. Insufficient emphasis on specific criteria for an acceptable match.
- **What are you going to do about it?** Implement a strict protocol for IGRT performed by students, requiring explicit, documented verification and approval of the alignment and shifts by a qualified supervising therapist *before* shifts are applied or before beam-on. Enhance training for both students and supervisors on common IGRT pitfalls and robust verification techniques. Review IGRT approval tolerances.

3. **Event 2: Incorrect Treatment Delivered**
    - Treatment is delivered to the patient with the incorrect alignment.
    - **What happened?** The patient received radiation to a position that was geographically misaligned from the intended target.

- **Why did it happen?** This was a direct consequence of the incorrect alignment performed in Event 1 not being detected by pre-treatment checks involving the student and supervising therapists.
- **What are you going to do about it?** Focus on preventing alignment errors and ensuring their detection *before* treatment delivery through improved IGRT protocols and supervisory practices.

4. **Event 3: Error Discovery by MD During Post-Treatment Image Review**
    - As usual, the MD reviews the treatment images *after* the treatment has been delivered.
    - The MD notices that the patient's alignment was off for that fraction.
    - The MD calls the (unspecified individual, likely therapy manager or physicist) regarding the improper alignment.
    - **What happened?** The misalignment was detected, but only after the incorrect treatment had been completed for that fraction.
    - **Why did it happen? (1st Why)** The MD has a routine process of reviewing IGRT images. Their clinical expertise and knowledge of the patient's anatomy and plan allowed them to identify the discrepancy that was missed pre-treatment.
    - **Why did it happen? (2nd Why - Deeper system/process cause)** The institutional process allows for MD image review to occur post-treatment rather than pre-treatment for every fraction (which is common for many, but not all, IGRT workflows, especially non-SBRT). While post-treatment review is valuable for quality assurance and identifying trends, it's too late to prevent an error for *that specific fraction*. The pre-treatment checks by therapy/physics (if physics was involved pre-tx) were not sufficient.
    - **Why did it happen? (3rd Why - Organizational/cultural cause)** A system that relies on post-treatment physician review as a primary method for catching daily setup errors, rather than ensuring robust real-time, pre-treatment verification by the treatment delivery team (therapists, and sometimes physicists for complex cases).
    - **What are you going to do about it?** While maintaining valuable post-treatment MD review, significantly strengthen the *pre-treatment* IGRT image review and approval process by therapists and, where appropriate,

physicists. Define clear objective criteria for acceptable alignment. Consider pre-treatment MD review for initial fractions or complex cases if not already standard.

5. **Event 4: Confirmation of Misalignment and Impact Analysis**
    - The Manager looks at the images and concurs that the alignment was off.
    - The MD and Manager ask physicists and dosimetrists to perform an analysis to determine the effect of the misalignment.
    - Per the analysis, the misaligned treatment had minimal effects on the dose delivered and the patient.
    - **What happened?** The error was confirmed, and its dosimetric and clinical impact were assessed.
    - **Why did it happen?** Standard and appropriate response after identification of a treatment delivery error.
    - **What are you going to do about it?** Ensure this process of analysis and impact assessment is consistently applied to all detected treatment errors.
6. **Event 5: Decision to Continue Treatment**
    - The MD reviewed the (new) plan/analysis and determined the clinic could proceed as normal with the original treatment plan for subsequent fractions.
    - **What happened?** A clinical decision was made regarding the patient's continued care based on the minimal impact of the error.
    - **Why did it happen?** The dosimetric analysis showed that the single misaligned fraction did not significantly compromise the overall treatment intent.
    - **What are you going to do about it?** Ensure that decisions to continue treatment after an error are well-documented, including the rationale and any patient communication.

---

### 2. Cause and Effect Diagramming

Identifying the conditions that resulted in the adverse event (misaligned treatment delivery).

**Categories of Potential Causes:**

- **People:**
  - *Student Therapist:*
    - (Potentially) Made an error in image registration, interpretation, shift calculation, or application.
    - (Potentially) Inexperience or lack of familiarity with specific anatomy/case.
  - *Supervising Therapist(s):*
    - Insufficient supervision or verification of student's IGRT alignment.
    - Failed to detect the misalignment before approving treatment.
    - (Potentially) Distracted, complacent, or over-reliant on student's capabilities or automated tools.
  - *MD (initially):* Not involved in pre-treatment alignment approval for this fraction (which is a common workflow but a factor in *when* the error was caught).
  - *Physicist (initially):* (Role in pre-treatment IGRT approval not specified, but if involved, also missed it).
- **Processes/Procedures:**
  - *Student Supervision Protocol for IGRT:*
    - May lack specific, stringent requirements for direct supervising therapist verification and sign-off of student-performed alignments *before* beam-on.
    - Unclear definition of what constitutes adequate supervision for IGRT.
  - *IGRT Alignment and Verification Protocol (General):*
    - Existing procedure used by therapists (and student) was not robust enough to prevent or detect the specific type of misalignment that occurred.
    - May lack objective criteria or detailed checklists for approving IGRT matches.

- May not adequately address common pitfalls in specific anatomical sites.
    - *Pre-Treatment Image Review Process:* The review by therapists (and student) before beam-on was insufficient to catch the error. MD review was post-treatment.
    - *Communication:* (Internal to therapy team during IGRT) Potential miscommunication or lack of challenge between student and supervisor.
- **Technology/Equipment (IGRT System):**
    - *Image Registration Software (Auto or Manual):* Potential for software limitations, user error in operating the tools, or misinterpretation of results.
    - *Image Quality (CBCT/Sim CT):* (Potential) Suboptimal image quality could make accurate registration or visual verification difficult.
    - *User Interface:* May not provide clear enough cues or warnings for potential misalignments.
- **Organizational/Environmental Factors:**
    - *Training Program (Students & Supervisors):*
        - (Potentially) Insufficient training for students on specific IGRT techniques or anatomical sites.
        - (Potentially) Insufficient training for supervising therapists on effective supervision methods for IGRT and common student errors.
    - *Workload/Time Pressure:* (Potential) Rushing could lead to less meticulous IGRT checks by students or supervisors.
    - *Safety Culture:*
        - (Potential) Culture may not sufficiently emphasize independent, critical verification by supervisors when students are performing tasks.
        - Post-treatment MD review as the primary catch for daily setup errors rather than robust pre-treatment therapist checks.
    - *Clarity of Roles and Responsibilities:* (Potential) Ambiguity in the level of scrutiny expected from the supervising therapist versus the student.

---

**3. Causal Statements, Actions, and Outcomes**

**Root Cause(s):**

1. **RC1: Inadequate Pre-Treatment Image-Guided Alignment Verification Process by the Treatment Delivery Team (Student and Supervising Therapists).** The established procedures and practices for verifying IGRT alignment by the student therapist and the supervising qualified therapist(s) before treatment delivery were insufficient to detect and correct the specific misalignment that occurred. This points to a failure in the human review and approval of the image registration and applied shifts.
    - **Contributing Factor:** Potential error in image registration, interpretation, or shift application by the student therapist.
    - **Contributing Factor:** Insufficiently rigorous or inadequately performed verification of the student's alignment by the supervising therapist(s) prior to beam-on.
    - **Contributing Factor:** Lack of clearly defined, objective, and consistently applied criteria for an acceptable IGRT match that all staff (including students and supervisors) must adhere to.
2. **RC2: Deficient Student Supervision Protocol for Critical IGRT Tasks.** The protocol guiding the supervision of student therapists performing IGRT did not ensure a level of direct oversight and mandatory, independent verification by the qualified supervising therapist that was adequate to prevent the misaligned treatment.
    - **Contributing Factor:** Potential ambiguity in the defined responsibilities and required level of direct intervention by the supervising therapist when a student is performing IGRT alignment.

**Key Factor Enabling Late Error Detection (Not a Root Cause of Error, but a Cause of Success in Discovery):**

- **Routine Post-Treatment Image Review by Physician:** The physician's standard practice of reviewing IGRT images after treatment delivery, although too late to prevent the error for that fraction, served as a safety net to detect the misalignment.

**Contributing Factor(s) (to the initial error or missed detection):**

- **CF1: Specific nature of the misalignment:** (The exact nature – e.g., wrong vertebral body, subtle rotation – is unknown from the summary, but some misalignments are harder to detect than others without meticulous, structured review).
- **CF2: Reliance on individual vigilance without sufficient system-level safeguards or detailed procedural prompts during the pre-treatment IGRT approval by therapists.**

**Actions and Outcome Measures:**

1. **Action (Addressing RC1 & RC2):** Develop and Implement a Strict, Standardized IGRT Alignment Verification Protocol, Especially When Students are Involved.
    - **Specifics:**
        - Mandate that for *all* IGRT alignments, and *especially* when a student is performing or assisting with the alignment:
            - A qualified supervising therapist must independently perform or meticulously re-verify every step of the image registration, target alignment, and shift application.
            - This verification must include comparing against defined anatomical landmarks, using all available imaging planes, and adhering to objective departmental tolerance criteria.
            - Documented sign-off by the supervising therapist confirming satisfactory alignment is required *before* beam-on.
        - Develop a clear checklist for IGRT alignment approval by therapists, outlining specific criteria and landmarks to verify for common treatment sites.
    - **Outcome Measure:** 100% documented supervising therapist verification and approval of IGRT alignments involving students. Reduction to zero of misaligned treatments. Audits of IGRT verification documentation.
2. **Action (Addressing RC1):** Enhance IGRT Training and Competency Assessment.
    - **Specifics:**
        - Provide comprehensive initial and recurrent IGRT training for all therapists and students, covering: common alignment errors, site-specific verification techniques, use of IGRT software tools,

understanding of departmental tolerances, and the importance of meticulous review.

- Implement practical competency assessments for IGRT alignment for both qualified therapists and students under supervision.
- Specifically train supervising therapists on effective methods for direct supervision and verification of student-performed IGRT.

- **Outcome Measure:** Documented completion of enhanced IGRT training and competency assessments. Improved consistency and accuracy in IGRT alignments.

3. **Action (Leveraging Successful Detection - Strengthening & Shifting Left):** Review and Optimize Timing and Process of Image Reviews.
   - **Specifics:**
     - While post-treatment MD review is valuable, explore opportunities for more real-time or proactive reviews for critical aspects or initial fractions, especially for complex cases or when students are heavily involved.
     - Strengthen the expectation and tools for therapists to perform a rigorous self-sufficient "final check" of alignment that meets objective criteria before proceeding, aiming to reduce reliance on post-treatment physician review as the primary catch for daily setup errors.
     - Define clear communication channels and triggers for therapists to request immediate physics or physician assistance if there is any uncertainty or difficulty in achieving satisfactory IGRT alignment *before* treatment.
   - **Outcome Measure:** Improved pre-treatment detection of potential alignment issues by the therapy team. Clear protocols for escalating IGRT uncertainties.
4. **Action (General IGRT Practice):** Define and Utilize Objective IGRT Approval Criteria.
   - **Specifics:** Establish and disseminate clear, site-specific, objective tolerance criteria for acceptable IGRT alignment (e.g., +/- X mm translation, +/- Y degrees rotation, required landmark congruence). These criteria should be used by all staff performing and verifying IGRT.

- **Outcome Measure:** Consistent application of objective IGRT approval criteria across all treatments.

5. **Action (Student Program):** Review and Refine Student Supervision Policies.
    - **Specifics:** Review the current policies for student participation in IGRT. Ensure they clearly define the levels of supervision required for different IGRT tasks, the exact responsibilities of the supervising therapist for verification and sign-off, and the student's scope of practice.
    - **Outcome Measure:** Updated and clear student supervision policies for IGRT.

---

**Lessons Learned from this Incident**

- **Supervision of Students Requires Active Verification:** When students perform critical tasks like IGRT, passive observation by supervising therapists is insufficient. Active, independent verification and formal sign-off by the qualified supervisor are essential before proceeding.
- **Pre-Treatment IGRT Approval by Therapy Team is a Critical Safety Step:** The final check and approval of IGRT alignment by the treating therapists (including supervisors when students are involved) before beam-on is a vital opportunity to catch errors. This check must be robust and meticulous.
- **Post-Treatment Review is Late for Error Prevention:** While valuable for QA and detecting trends, physician review of IGRT images *after* treatment delivery cannot prevent an error for that specific fraction. Emphasis must be on robust *pre-treatment* verification.
- **Human Factors in Image Interpretation:** Image registration and interpretation are susceptible to human error, even with advanced software. Clear protocols, objective criteria, and independent checks are needed to mitigate this.
- **Importance of Holistic Anatomical Review:** (Although not explicitly stated as the method of MD detection, it's often key) Looking at the broader anatomical context, not just the immediate target or bony match, can help identify subtle but significant misalignments.
- **Minimal Dosimetric Impact Doesn't Negate Process Failure:** Even if a specific error results in minimal dosimetric consequence, the underlying process failures that allowed the error to occur must be addressed to prevent future, potentially more significant, errors.

- **Documentation of IGRT:** (Implicit) Clear documentation of who performed and verified IGRT, the shifts applied, and images reviewed is important for accountability and error investigation.